\documentclass[eqsecnum,preprint,showpacs,amsmath,aps,nofootinbib]
  {revtex4}  
\usepackage{epsfig}
\usepackage{amssymb}
\usepackage[dvips]{color}
\usepackage[latin1]{inputenc}
\usepackage{graphicx}

\makeatletter
\renewcommand*{\@fnsymbol}[1]{\ensuremath{\ifcase#1\or *\or \dagger\or
    \ddagger\or 
   \mathsection\or **\or \dagger\dagger
   \or \ddagger\ddagger \else\@ctrerr\fi}}
\makeatother

\begin{document}
\title{Self-dual road to noncommutative gravity
with twist: a new analysis}

\author{Elisabetta Di Grezia}
\email[E-mail: ]{digrezia@na.infn.it}
\affiliation{Istituto Nazionale di Fisica Nucleare, Sezione di
Napoli, Complesso Universitario di Monte S. Angelo, 
Via Cintia Edificio 6, 80126 Napoli, Italy}

\author{Giampiero Esposito}
\email[E-mail: ]{gesposit@na.infn.it}
\affiliation{Istituto Nazionale di Fisica Nucleare, Sezione di
Napoli, Complesso Universitario di Monte S. Angelo, 
Via Cintia Edificio 6, 80126 Napoli, Italy}

\author{Patrizia Vitale} 
\email[E-mail: ]{patrizia.vitale@na.infn.it}
\affiliation{Universit\`a di Napoli Federico II, 
Dipartimento di Fisica, Complesso Universitario di Monte S. Angelo,
Via Cintia Edificio 6, 80126 Napoli, Italy\\
Istituto Nazionale di Fisica Nucleare, Sezione di
Napoli, Complesso Universitario di Monte S. Angelo,
Via Cintia Edificio 6, 80126 Napoli, Italy}

\date{\today}

\begin{abstract}
The field equations of noncommutative gravity can be obtained
by replacing all exterior products by twist-deformed exterior products 
in the action functional of general relativity,  
and are here studied by requiring
that the torsion $2$-form should vanish, and that the 
Lorentz-Lie-algebra-valued part of the full connection $1$-form 
should be self-dual. Other two conditions, expressing 
self-duality of a pair $2$-forms occurring in the full curvature
$2$-form, are also imposed. This leads to a systematic solution strategy, 
here displayed for the first time, where all parts of the connection
$1$-form are first evaluated, hence the full curvature $2$-form, and
eventually all parts of the tetrad $1$-form, when expanded on the basis 
of $\gamma$-matrices. By assuming asymptotic expansions which hold 
up to first order in the noncommutativity matrix in the
neighbourhood of the vanishing value for noncommutativity, we find a
family of self-dual solutions of the field equations. This is generated
by solving first a inhomogeneous wave equation on $1$-forms in
a classical curved spacetime (which is itself self-dual and solves
the vacuum Einstein equations), subject to the Lorenz gauge condition.
In particular, when the classical undeformed geometry is Kasner 
spacetime, the above scheme is fully computable out of solutions of
the scalar wave equation in such a Kasner model. 
\end{abstract}


\maketitle

\section{Introduction}

The tetrad formalism of Cartan \cite{Cartan} has proved useful both
in gravity (see e.g. Sec. 10.1 of Ref. \cite{DeWi83}) and
supergravity \cite{SUGRA76}, as well as in gravitational
instanton theory \cite{Eguchi} and in modern approaches to the
Hamiltonian formulation of general relativity \cite{Ash87, Capo91}.
Within this framework, one assumes that a set of local Lorentz frames
exist, whose global existence is ensured if the classical spacetime
manifold is parallelizable. The covariant components $g_{\mu \nu}$ of 
the metric tensor can then be re-expressed through tetrad covectors
$e_{\; \mu}^{a}$ in the form 
\begin{equation}
g_{\mu \nu}=e_{\; \mu}^{a} \; e_{\; \nu}^{b} \eta_{ab}
=e_{b \mu} \; e_{\; \nu}^{b},
\end{equation}
so that, on defining the tetrad $1$-forms 
\begin{equation}
e^{a} \equiv e_{\; \mu}^{a} {\rm d}x^{\mu},
\end{equation}
the spacetime metric reads eventually
\begin{equation}
g=g_{\mu \nu}{\rm d}x^{\mu} \otimes {\rm d}x^{\nu}
=e^{a} \otimes e^{b} \eta_{ab}.
\end{equation}
Of course, the `dual' description in terms of tetrad vectors 
$e_{\; a}^{\mu}$ is also possible. One then finds, from the
contravariant metric components
\begin{equation}
g^{\mu \nu}=e_{\; a}^{\mu} \; e_{\; b}^{\nu} \eta^{ab}
=e^{\mu b} \; e_{\; b}^{\nu},
\end{equation}
jointly with the tetrad vector fields \cite{DeWi83}
\begin{equation}
e_{a} \equiv e_{\; a}^{\mu} {\partial \over \partial x^{\mu}},
\end{equation}
the other useful formula
\begin{equation}
g=g^{\mu \nu}{\partial \over \partial x^{\mu}} \otimes
{\partial \over \partial x^{\nu}}
=e_{a} \otimes e_{b} \eta^{ab}.
\end{equation}
On the other hand, several investigations of noncommutative gravity
have exploited the tetrad and spin-connection as well, but replacing the
ordinary exterior products of forms with deformed exterior products
\cite{AC09}. This would be fair enough, with no need for extra
mathematical machinery, if it were not for the fact that attempts of
providing a rigorous definition of noncommutative gravity equations
{\it jointly with their solution} had been unsuccessful, at least
within the framework of twist differential geometry (see appendix A)
in the version considered in Refs. \cite{ADE11, DEFV13}. 
One of the basic aspects of tetrad 
formalism for noncommutative gravity is to expand the tetrad on the
basis of $\gamma$-matrices, so that one deals actually with the space
of tetrad $1$-forms with components given by $4 \times 4$ matrices, 
i.e. (see appendix B)
\begin{equation}
V_{j}^{\; k}=(V_{\mu})_{j}^{\; k}{\rm d} x^{\mu},
\end{equation}
where 
\begin{equation}
(V_{\mu})_{j}^{\; k}=V_{\; \mu}^{a}(\gamma_{a})_{j}^{\; k}
+{\widetilde V}_{\; \mu}^{a}(\gamma_{a}\gamma_{5})_{j}^{\; k}.
\end{equation}
Hereafter, following Ref. \cite{AC09}, the noncommutativity we
consider is given by the Moyal-Weyl $\star$-product associated
with a constant antisymmetric matrix $\theta^{\rho \sigma}$ in
the generic coordinates $x^{\mu}$, which obey the `deformed'
commutation law
\begin{equation}
x^{\rho} \star x^{\sigma}-x^{\sigma} \star x^{\rho}
\equiv {\rm i}\theta^{\rho \sigma}.
\end{equation}
Moreover, following again Ref. \cite{AC09}, the additive structures
are not modified, while multiplicative structures get deformed. 
Thus, {\it the notion of tensors that we consider remains the same 
as in the classical commutative setting, while the
tensor product and the exterior product (and only these structures)
are deformed}. For a deeper look at the mathematical foundations,
we refer the reader to appendix A and to the work in Ref. \cite{AS12}.

Both $V_{\; \mu}^{a}$ and ${\widetilde V}_{\; \mu}^{a}$ depend on the
noncommutativity matrix $\theta^{\rho \sigma}=\theta^{[\rho \sigma]}$,
and are approximated by means of
even and odd \cite{AC09} asymptotic expansions 
in the neighborhood of $\theta^{\rho \sigma}=0$ of the form
\begin{equation}
V_{\; \mu}^{a}(\theta)=V_{\; \mu}^{a}(-\theta) \sim
e_{\; \mu}^{a}+{\rm O}(\theta^{2}),
\end{equation}
\begin{equation}
{\widetilde V}_{\; \mu}^{a}(\theta)=-{\widetilde V}_{\; \mu}^{a}(-\theta)
\sim \theta^{\rho \sigma}P_{\; \mu [\rho \sigma]}^{a}
+{\rm O}(\theta^{3}).
\end{equation}
(Notice however that ${\widetilde V}$ has no commutative analogue, 
it is generated by the requirement that the action functional introduced 
in next section be fully invariant under $\star-$gauge transformations. 
We refer to the appendix B for details).

This is indeed a crucial point of all our analysis. The full theory,
at nonperturbative level, is nonlocal, but there is not yet any
experimental evidence of finite effects resulting from finite
values of $\theta^{\rho \sigma}$. Thus, we limit ourselves to
studying the behavior of noncommutative gravity in the
neighborhood of $\theta^{\rho \sigma}=0$. The existence of
even and odd parts of the tetrad is an exact property \cite{AC09},
but we focus on their asymptotics (1.10) and (1.11).  

The full connection $1$-form is also a $4 \times 4$ matrix of
$1$-forms \cite{AC09} expandable as (see appendix B)
\begin{equation}
\Omega_{j}^{\; k}=(\Omega_{\mu})_{j}^{\; k}{\rm d} x^{\mu}
=\left[\omega_{\mu}^{ab}(\Gamma_{ab})_{j}^{\; k}
+{\rm i}\omega_{\mu}\delta_{j}^{\; k}
+{\widetilde \omega}_{\mu}(\gamma_{5})_{j}^{\; k}\right]
{\rm d} x^{\mu},
\end{equation}
with
\begin{equation}
\Gamma_{ab}={1\over 4}\gamma_{ab}={1\over 8}(\gamma_{a}\gamma_{b}
-\gamma_{b}\gamma_{a})={1\over 4}\gamma_{[a}\gamma_{b]}.
\end{equation}
The components $\omega_{\mu}^{ab}$ take values
in the Lie algebra of the Lorentz group, and hence carry
Lorentz-frame indices. One deals therefore with a
$1$-form $\omega^{ab}=\omega_{\mu}^{ab}{\rm d} x^{\mu}$, which yields 
the usual spin-connection in the commutative limit, 
while $\omega_{\mu}$ and
${\widetilde \omega}_{\mu}$ are components of purely noncommutative  
$1$-forms $\omega=\omega_{\mu}{\rm d}x^{\mu}$ and 
${\widetilde \omega}={\widetilde \omega}_{\mu}{\rm d}x^{\mu}$,  
introduced, as ${\widetilde V}$, to fulfill the 
$\star$-gauge invariance of the theory (see again appendix B). 

The full curvature $2$-form is defined by
\begin{equation}
{\cal R} \equiv {\rm d} \Omega-\Omega \wedge_{\star} \Omega,
\end{equation}
and, by writing explicitly all matrix, coordinate and Lorentz-frame
indices, reads as
\begin{equation}
{\cal R}_{j}^{\; k}={1\over 2}({\cal R}_{\mu \nu})_{j}^{\; k}
{\rm d} x^{\mu} \wedge {\rm d} x^{\nu},
\end{equation}
where \cite{AC09} 
\begin{equation}
({\cal R}_{\mu \nu})_{j}^{\; k}=R_{\mu \nu}^{ab}
(\Gamma_{ab})_{j}^{\; k}+{\rm i}r_{\mu \nu}\delta_{j}^{\; k}
+{\widetilde r}_{\mu \nu}(\gamma_{5})_{j}^{\; k}.
\end{equation}
With this notation, $R_{\mu \nu}^{ab}$ are the components of the
Lorentz-Lie-algebra-valued $2$-form (for a more precise language,
see what we write after Eq. (B4) of appendix B) 
$$
R^{ab}={1\over 2}R_{\mu \nu}^{ab}
{\rm d}x^{\mu} \wedge {\rm d}x^{\nu},
$$
while $r_{\mu \nu}$ and ${\widetilde r}_{\mu \nu}$ are the components
of the $2$-forms
$$
r={1\over 2}r_{\mu \nu}{\rm d}x^{\mu} \wedge {\rm d}x^{\nu} \; 
{\rm and} \;
{\widetilde r}={1\over 2}{\widetilde r}_{\mu \nu}
{\rm d}x^{\mu} \wedge {\rm d}x^{\nu}.
$$
By virtue of (1.12) and (1.14)-(1.16), and exploiting the definition of
Hodge dual (both $\omega^{ab}$ and $R^{ab}$ can be treated as 
$2$-forms with respect to Lorentz-frame indices, as discussed
in appendix C)
\begin{equation}
{ }^{(*)}\omega^{ab} \equiv {1\over 2}\varepsilon_{\; \; \; cd}^{ab}
\omega^{cd},
\end{equation}
\begin{equation}
{ }^{(*)}R^{ab} \equiv {1\over 2}\varepsilon_{\; \; \; cd}^{ab}R^{cd},
\end{equation}
and the twist-deformed exterior product of $1$-forms 
$\alpha_{\mu}{\rm d}x^{\mu}$
and $\beta_{\nu}{\rm d}x^{\nu}$,
\begin{equation}
\alpha \wedge_{\star} \beta=\alpha_{[\mu} \star \beta_{\nu]}
{\rm d}x^{\mu} \wedge {\rm d}x^{\nu},
\end{equation}
with 
\begin{equation}
\alpha_{\mu} \star \beta_{\nu} \sim \alpha_{\mu} \beta_{\nu}
+{{\rm i}\over 2}\theta^{\rho \sigma}(\partial_{\rho}\alpha_{\mu})
(\partial_{\sigma}\beta_{\nu})+{\rm O}(\theta^{2})
\; {\rm as} \; \theta^{\rho \sigma} \rightarrow 0,
\end{equation}
\begin{equation}
\alpha_{[\mu} \star \beta_{\nu]} \equiv 
{1\over 2} (\alpha_{\mu} \star \beta_{\nu}
-\alpha_{\nu} \star \beta_{\mu}),
\end{equation}
one finds eventually the components (cf. \cite{AC09})
\begin{eqnarray}
R_{\mu \nu}^{ab}&=& 2 \partial_{[\mu} \omega_{\nu ]}^{ab}
+\biggr(\omega_{\; c [\mu}^{b} \star \omega_{\nu]}^{ca}
-\omega_{\; c[\mu}^{a} \star \omega_{\nu]}^{cb}\biggr)
-2{\rm i}\Bigr(\omega_{[\mu}^{ab} \star \omega_{\nu ]}
+\omega_{[\mu} \star \omega_{\nu ]}^{ab}\Bigr)
\nonumber \\
&-& 2{\rm i}\Bigr[{ }^{(*)}\omega_{[\mu}^{ab}
\star {\widetilde \omega}_{\nu ]}
+{\widetilde \omega}_{[\mu} \star { }^{(*)}\omega_{\nu ]}^{ab}\Bigr],
\end{eqnarray}
\begin{equation}
r_{\mu \nu}=2\partial_{[\mu}\omega_{\nu]}-{{\rm i}\over 4}
\omega_{cd [\mu} \star \omega_{\nu ]}^{cd}
-2{\rm i}\Bigr(\omega_{[\mu} \star \omega_{\nu ]}
-{\widetilde \omega}_{[\mu} \star {\widetilde \omega}_{\nu ]}\Bigr),
\end{equation}
\begin{equation}
{\widetilde r}_{\mu \nu}=2\partial_{[\mu} {\widetilde \omega}_{\nu]}
+{{\rm i}\over 4}{ }^{(*)}\omega_{cd [\mu} \star 
\omega_{\nu]}^{cd}-2{\rm i}\Bigr(\omega_{[\mu} \star
{\widetilde \omega}_{\nu ]}+{\widetilde \omega}_{[\mu} \star
\omega_{\nu ]}\Bigr).
\end{equation}
The use of Hodge duals is suggested by what one finds in simpler
circumstances. For example, in general relativity, self-duality
(resp. anti-self-duality) of the spin-connection $1$-form, i.e.
\begin{equation}
{ }^{(*)}\omega^{ab}=\pm {\rm i} \omega^{ab},
\end{equation}
is a sufficient condition for self-duality (resp. anti-self-duality)
of the curvature $2$-form:
\begin{equation}
{ }^{(*)}R^{ab}=\pm {\rm i}R^{ab}.
\end{equation}
More precisely, Eq. (1.26) is important because,
from the condition of vanishing torsion $2$-form, i.e.
\begin{equation}
T \equiv {\rm d}e-\omega \wedge e =0,
\end{equation}
one finds a solution of the vacuum Einstein equations, since the
latter read as
\begin{equation}
{ }^{(*)}R \wedge e=0,
\end{equation}
while from ${\rm d}T=0$ and from $R={\rm d} \omega -\omega \wedge \omega$ 
and associativity of the exterior product one finds
\begin{equation}
R \wedge e=0,
\end{equation}
which coincides with Eq. (1.28) upon imposing self-duality or
anti-self-duality: ${ }^{(*)}R=\pm {\rm i}R$ according to Eq. (1.26).

The plan of our paper is hence as follows. In Sec. II we write the
torsion-free field equations of noncommutative gravity, without any
coupling to other fields, and in Sec. III we consider their self-dual
form. In Sec. IV we study the self-duality conditions on $\omega^{ab}$, 
while Sec. V expresses self-duality of the
$2$-forms $r$ and ${\widetilde r}$. The remaining self-dual equations
are re-expressed in Sec. VI. In Secs. from VII to XI 
we study, to first order in 
noncommutativity, the resulting set of equations for the $1$-form
$\omega^{ab}$, jointly with the $1$-forms $\omega,{\widetilde \omega}$
and the components $V_{\mu}^{a}$ and ${\widetilde V}_{\mu}^{a}$ of the
tetrad $1$-form. Our results and the open problems are described in
section XII, while appendices A, B and C describe in detail the
foundations of the concepts we have been using and the operations
we have been performing.

\section{Torsion-free field equations}\label{section2}

The basic assumption of quantum theory \cite{DW65} is that every 
isolated dynamical system can be described by a characteristic
action functional $S$. 
Our paper does not deal with quantum theory, but prepares the ground 
for it by studying the Euler-Lagrange equations for a given choice of
action functional. Relying upon Ref. \cite{AC09}, the starting point is
an action for gravity where all exterior products 
are replaced by twist-deformed exterior products (see Appendix A for 
a definition), i.e.
\begin{equation}
S=\int {\rm Tr} \Bigr({\rm i}{\cal R} \wedge_{\star} V \wedge_{\star}
V \gamma_{5}\Bigr). \label{action}
\end{equation}
This action is invariant under ordinary diffeomorphisms as well as 
$\star$-diffeomorphisms (we refer to Sec. 3 of Appendix A for 
details), and it is also invariant under $\star$-gauge transformations, 
as described in appendix B.  It leads to the field equations
\begin{equation}
{\rm Tr} \Bigr[\gamma_{c}\gamma_{5}(V \wedge_{\star}{\cal R}
+{\cal R} \wedge_{\star}V)\Bigr]=0,
\end{equation}
\begin{equation}
{\rm Tr} \Bigr[\gamma_{c}(V \wedge_{\star}{\cal R}
+{\cal R} \wedge_{\star}V)\Bigr]=0.
\end{equation}
Bearing in mind what we said in the Introduction, we now consider the
torsion $2$-form in the noncommutative setting \cite{AC09}, i.e.
\begin{equation}
T \equiv {\rm d}V-\Omega \wedge_{\star} V - V \wedge_{\star} \Omega,
\end{equation} 
and investigate the consequence of requiring it should vanish. This
is not mandatory but certainly legitimate. Indeed, from $T=0$ one finds
\begin{equation}
{\rm d}V=\Omega \wedge_{\star} V + V \wedge_{\star} \Omega,
\end{equation}
while from ${\rm d}T=0$ one obtains
\begin{eqnarray}
0&=& -{\rm d}(\Omega \wedge_{\star} V)
-{\rm d}(V \wedge_{\star} \Omega)
\nonumber \\
&=& -({\rm d} \Omega)\wedge_{\star}V +\Omega \wedge_{\star} {\rm d}V
-{\rm d}V \wedge_{\star} \Omega +V \wedge_{\star} {\rm d} \Omega
\nonumber \\
&=& -({\cal R}+\Omega \wedge_{\star} \Omega)\wedge_{\star}V
+\Omega \wedge_{\star}(\Omega \wedge_{\star} V 
+V \wedge_{\star} \Omega) 
\nonumber \\
&-& (\Omega \wedge_{\star}V + V \wedge_{\star} \Omega)
\wedge_{\star}\Omega + V \wedge_{\star}({\cal R}+
\Omega \wedge_{\star} \Omega)
\nonumber \\
&=& -{\cal R} \wedge_{\star} V + V \wedge_{\star} {\cal R},
\end{eqnarray}
because also the twist-deformed exterior product is associative 
\cite{AC09}. Thus, we find the simple but nontrivial property
$$
{\cal R} \wedge_{\star} V=V \wedge_{\star} {\cal R},
$$
and the torsion-free version of the field equations (2.2) and (2.3)
becomes
\begin{equation}
{\rm Tr}\Bigr[\gamma_{c}\gamma_{5}V \wedge_{\star} {\cal R}
\Bigr]=0,
\end{equation}
\begin{equation}
{\rm Tr}\Bigr[\gamma_{c} V \wedge_{\star} {\cal R} \Bigr]=0.
\end{equation}
An equivalent result would be obtained by defining the torsion 
$2$-form according to
$$
T \equiv {\rm d}V-\Omega \wedge_{\star} V,
$$
because Eq. (2.6) would then reduce to ${\cal R} \wedge_{\star}V=0$,
which again turns Eqs. (2.2) and (2.3) into the torsion-free form
(2.7) and (2.8).

From the expansion (1.8) we note that
\begin{equation}
\gamma_{5}V=-{\widetilde V}^{a}\gamma_{a}-V^{a}\gamma_{a}\gamma_{5},
\end{equation}
and hence it is a priori clear that the $2$ sets of equations are
obtained one from the other by interchanging $V_{\mu}^{a}$ with
${\widetilde V}_{\mu}^{a}$, as was pointed out at a later stage in
Ref. \cite{ADE11}. Now, from the decompositions (1.8) and (1.16), the
traces in Eqs. (2.7) and (2.8) are found to be (see appendix B)
\begin{eqnarray}
0 &=& {\rm Tr}\Bigr[\gamma_{c}\gamma_{5}{\rm i} V \wedge_{\star}
{\cal R}\Bigr] \nonumber \\
&=& -\varepsilon_{abcd}V^{d} \wedge_{\star}R^{ab}
-4{\rm i}\eta_{cd}V^{d} \wedge_{\star} {\widetilde r}
-{\rm i}(\eta_{bc}\eta_{ad}-\eta_{ac}\eta_{bd})
{\widetilde V}^{d} \wedge_{\star} R^{ab} \nonumber \\
&+& 4 \eta_{cd}{\widetilde V}^{d} \wedge_{\star}r,
\end{eqnarray}
\begin{eqnarray}
0&=& {\rm Tr}[\gamma_{c} {\rm i} V \wedge_{\star} {\cal R}]
\nonumber \\
&=& \varepsilon_{abcd}{\widetilde V}^{d} \wedge_{\star}R^{ab}
+4{\rm i}\eta_{cd}{\widetilde V}^{d} \wedge_{\star} {\widetilde r}
+{\rm i}(\eta_{bc}\eta_{ad}-\eta_{ac}\eta_{bd})
V^{d} \wedge_{\star} R^{ab} \nonumber \\
&-& 4 \eta_{cd}V^{d} \wedge_{\star}r.
\end{eqnarray}
Now we exploit the definition (1.18) of Hodge dual, jointly with
the identity
$$
(\eta_{ad}\eta_{bc}-\eta_{ac}\eta_{bd})R^{ab}=R_{dc}-R_{cd}=2R_{dc},
$$
to write Eqs. (2.7) and (2.8) in the form
\begin{equation}
V^{d} \wedge_{\star} { }^{(*)}R_{cd}
+{\widetilde V}^{d} \wedge_{\star}(-{\rm i}R_{cd})
+2 \Bigr({\rm i}V_{c} \wedge_{\star}{\widetilde r}
-{\widetilde V}_{c} \wedge_{\star}r \Bigr)=0,
\end{equation}
\begin{equation}
{\widetilde V}^{d} \wedge_{\star} { }^{(*)}R_{cd}
+V^{d} \wedge_{\star}(-{\rm i}R_{cd})
+2 \Bigr({\rm i}{\widetilde V}_{c} \wedge_{\star}{\widetilde r}
-V_{c} \wedge_{\star}r \Bigr)=0.
\end{equation}
Recall now that any $2$-form $F$ can be decomposed into its self-dual
($F^{+}$) and anti-self-dual part ($F^{-}$) according to
$F=F^{+}+F^{-}$, where (the imaginary unit occurs because of the
Lorentzian signature)
$$
F^{+} \equiv {1\over 2}(F-{\rm i}{ }^{(*)}F), \;
F^{-} \equiv {1\over 2}(F+{\rm i}{ }^{(*)}F).
$$
We apply this decomposition to the Lorentz-Lie-algebra-valued
$2$-form $R_{cd}$ in Eqs. (2.12) and (2.13), and also to the
$2$-forms $r$ and ${\widetilde r}$ therein. We further multiply both
equations by $-{\rm i}$, and hence get
\begin{equation}
V^{d} \wedge_{\star} (R_{cd}^{-}-R_{cd}^{+})
+{\widetilde V}^{d}\wedge_{\star}(R_{cd}^{-}+R_{cd}^{+})
-2 V_{c} \wedge_{\star}({\widetilde r}^{-}+{\widetilde r}^{+})
-2{\rm i}{\widetilde V}_{c} \wedge_{\star} (r^{-}+r^{+})=0,
\end{equation}
\begin{equation}
{\widetilde V}^{d} \wedge_{\star} (R_{cd}^{-}-R_{cd}^{+})
+V^{d}\wedge_{\star}(R_{cd}^{-}+R_{cd}^{+})
-2 {\widetilde V}_{c} \wedge_{\star}
({\widetilde r}^{-}+{\widetilde r}^{+})
-2{\rm i}V_{c} \wedge_{\star} (r^{-}+r^{+})=0.
\end{equation}

\section{Self-dual torsion-free equations}

In the self-dual case one sets to $0$ all anti-self-dual parts
of the curvature, i.e.
\begin{equation}
R_{cd}^{-}=0, \; r^{-}=0, \; {\widetilde r}^{-}=0,
\end{equation}
where the Hodge dual of $r$ and ${\widetilde r}$,
{\it provided one works to linear order in} $\theta$ as we are doing,
can be re-espressed through the undeformed 
Levi-Civita symbol with coordinate indices for curved spacetime, i.e.
(see detailed discussion in appendix C)
\begin{equation}
{ }^{(*)}r_{\mu \nu} \sim {1\over 2}
\varepsilon_{\mu \nu}^{\; \; \; \rho \sigma}r_{\rho \sigma}
+{\rm O}(\theta^{2}), \;
{ }^{(*)}{\widetilde r}_{\mu \nu} \sim {1\over 2}
\varepsilon_{\mu \nu}^{\; \; \; \rho \sigma}
{\widetilde r}_{\rho \sigma}
+{\rm O}(\theta^{2}), {\rm as} \; \theta \rightarrow 0.
\end{equation}
We now insert (3.1) into (2.14) and (2.15), and add up the $2$
resulting equations to find the set of `self-dual equations'
\begin{equation}
\Bigr({\widetilde V}^{d}-V^{d}\Bigr)\wedge_{\star}R_{cd}^{+}
-2 V_{c} \wedge_{\star} {\widetilde r}^{+}
-2{\rm i}{\widetilde V}_{c} \wedge_{\star} r^{+}=0,
\end{equation}
\begin{equation}
\Bigr({\widetilde V}_{c}+V_{c}\Bigr) \wedge_{\star}
\Bigr({\widetilde r}^{+}+{\rm i}r^{+}\Bigr)=0.
\end{equation}

In the anti-self-dual case, one sets instead to $0$ the self-dual
parts of the curvature, i.e.
\begin{equation}
R_{cd}^{+}=0, \; r^{+}=0, \; {\widetilde r}^{+}=0.
\end{equation}
By inserting (3.5) into (2.14) and (2.15), and subtracting the $2$
resulting equations, we find the `anti-self-dual equations'
\begin{equation}
\Bigr({\widetilde V}^{d}+V^{d}\Bigr)\wedge_{\star}R_{cd}^{-}
-2 V_{c} \wedge_{\star} {\widetilde r}^{-}
-2{\rm i}{\widetilde V}_{c} \wedge_{\star} r^{-}=0,
\end{equation}
\begin{equation}
\Bigr(V_{c}-{\widetilde V}_{c}\Bigr) \wedge_{\star}
\Bigr({\widetilde r}^{-}-{\rm i}r^{-}\Bigr)=0.
\end{equation}

\section{Self-duality of $\omega^{ab}$}

The conditions $R_{cd}^{-}=0$ or $R_{cd}^{+}=0$ considered in section 3
may still lead to complicated equations, as is clear from (1.22). 
However, Eq. (1.22) tells us something more helpful: if the
$1$-form $\omega^{ab}$ is self-dual or anti-self-dual, according
to (1.25), this is a sufficient condition for self-duality or
anti-self-duality of $R^{ab}$ itsef. Indeed one finds, by virtue of (1.25),
\begin{equation}
{ }^{(*)}\Bigr[{\rm d}\omega^{ab}
-{\rm i}(\omega^{ab}\wedge_{\star} \omega
+\omega \wedge_{\star} \omega^{ab})\Bigr]
=\pm {\rm i}\Bigr[{\rm d}\omega^{ab}
-{\rm i}(\omega^{ab}\wedge_{\star} \omega
+\omega \wedge_{\star}\omega^{ab})\Bigr],
\end{equation}
\begin{eqnarray}
\; & \; & { }^{(*)}\Bigr[-{\rm i}({ }^{(*)}\omega^{ab}
\wedge_{\star}{\widetilde \omega}+{\widetilde \omega}
\wedge_{\star} { }^{(*)}\omega^{ab})\Bigr]
=-\Bigr[-{\rm i}(\omega^{ab}\wedge_{\star} {\widetilde \omega}
+{\widetilde \omega} \wedge_{\star} \omega^{ab})\Bigr]
\nonumber \\
&=& \pm {\rm i} \Bigr[-{\rm i}({ }^{(*)}\omega^{ab}
\wedge_{\star} {\widetilde \omega}
+{\widetilde \omega} \wedge_{\star} 
{ }^{(*)}\omega^{ab})\Bigr].
\end{eqnarray}
Moreover, by virtue of the identity \cite{ADE11}
\begin{equation}
\varepsilon^{abcd}\varepsilon_{defg}=\delta_{d}^{a}
\Bigr(\delta_{e}^{b}\delta_{f}^{c}-\delta_{e}^{c}\delta_{f}^{b}\Bigr)
+\delta_{d}^{b}\Bigr(\delta_{e}^{c}\delta_{f}^{a}-\delta_{e}^{a}
\delta_{f}^{c}\Bigr) 
+ \delta_{d}^{c}\Bigr(\delta_{e}^{a}\delta_{f}^{b}
-\delta_{e}^{b}\delta_{f}^{a}\Bigr),
\end{equation}
one finds
\begin{eqnarray}
\; & \; & { }^{(*)}\Bigr(\omega_{\; c}^{a} \wedge_{\star} 
\omega^{cb} \Bigr)={1\over 2}\varepsilon_{\; \; \; ef}^{ab}
\omega_{\; c [\mu}^{e} \star \omega_{\nu ]}^{cf} 
{\rm d}x^{\mu} \wedge {\rm d}x^{\nu} 
=\mp {{\rm i}\over 4}\varepsilon_{\; \; \; ef}^{ab}
\varepsilon_{\; cpq}^{e} \omega_{[\mu}^{pq} \star \omega_{\nu ]}^{cf}
{\rm d}x^{\mu} \wedge {\rm d}x^{\nu} 
\nonumber \\
&=& \mp {{\rm i}\over 2}\Bigr(\omega_{[\mu}^{cb} \star 
\omega_{\; c \nu]}^{a}+\omega_{[\mu}^{ac} \star 
\omega_{\; c \nu]}^{b}\Bigr){\rm d}x^{\mu} \wedge {\rm d}x^{\nu}
=\pm {\rm i} \omega_{\; c[\mu}^{a} \star \omega_{\nu]}^{cb}
{\rm d}x^{\mu} \wedge {\rm d}x^{\nu},
\end{eqnarray}
and analogous procedure holds for the Hodge dual of 
$\omega_{\; c}^{b} \wedge_{\star} \omega^{ca}$. Hence one finds
\begin{eqnarray}
{ }^{(*)}R^{ab}&=& { }^{(*)}\Bigr[{\rm d}\omega^{ab}
-{1\over 2}\omega_{\; c}^{a} \wedge_{\star} \omega^{cb}
+{1\over 2}\omega_{\; c}^{b} \wedge_{\star} \omega^{ca} 
\nonumber \\
&-& {\rm i}(\omega^{ab}\wedge_{\star}\omega + \omega \wedge_{\star}
\omega^{ab})-{\rm i}({ }^{(*)}\omega^{ab}\wedge_{\star} 
{\widetilde \omega}+{\widetilde \omega} \wedge_{\star}
{ }^{(*)}\omega^{ab})\Bigr]
\nonumber \\
&=& \pm {\rm i} R^{ab},
\end{eqnarray}
provided that
\begin{equation}
\varepsilon_{\; \; \; cd}^{ab} \omega_{\mu}^{cd}
=2{\rm i} \omega_{\mu}^{ab}.
\end{equation}

\section{Self-duality of the $2$-forms $r$ and ${\widetilde r}$}

Self-duality of the $2$-forms $r$ and ${\widetilde r}$ means setting
to $0$ their anti-self-dual parts, which implies that
\begin{equation}
r_{\mu \nu}=-{\rm i}\varepsilon_{\mu \nu}^{\; \; \; \rho \sigma}
r_{\rho \sigma},
\end{equation}
\begin{equation}
{\widetilde r}_{\mu \nu}
=-{\rm i}\varepsilon_{\mu \nu}^{\; \; \; \rho \sigma}
{\widetilde r}_{\rho \sigma}.
\end{equation}
By virtue of (1.23)-(1.25), this leads to the equations
\begin{eqnarray}
\; & \; & 2 \partial_{[\mu} \omega_{\nu]}
+2{\rm i}\varepsilon_{\mu \nu}^{\; \; \; \rho \sigma}
\partial_{[\rho} \omega_{\sigma]}-{{\rm i}\over 4}
\omega_{cd [\mu} \star \omega_{\nu]}^{cd}
+{1\over 4}\varepsilon_{\mu \nu}^{\; \; \; \rho \sigma}
\omega_{cd [\rho} \star \omega_{\sigma]}^{cd}
\nonumber \\
&-& 2{\rm i}\Bigr(\omega_{[\mu} \star \omega_{\nu]}
-{\widetilde \omega}_{[\mu} \star {\widetilde \omega}_{\nu]}\Bigr)
+2 \varepsilon_{\mu \nu}^{\; \; \; \rho \sigma}
\Bigr(\omega_{[\rho} \star \omega_{\sigma]}
-{\widetilde \omega}_{[\rho} \star {\widetilde \omega}_{\sigma]}
\Bigr)=0,
\end{eqnarray}
\begin{eqnarray}
\; & \; & 2 \partial_{[\mu} {\widetilde \omega}_{\nu]}
+2{\rm i}\varepsilon_{\mu \nu}^{\; \; \; \rho \sigma}
\partial_{[\rho} {\widetilde \omega}_{\sigma]}-{1 \over 4}
\omega_{cd [\mu} \star \omega_{\nu]}^{cd}
-{{\rm i}\over 4}\varepsilon_{\mu \nu}^{\; \; \; \rho \sigma}
\omega_{cd [\rho} \star \omega_{\sigma]}^{cd}
\nonumber \\
&-& 2{\rm i}\Bigr(\omega_{[\mu} \star {\widetilde \omega}_{\nu]}
+{\widetilde \omega}_{[\mu} \star \omega_{\nu]}\Bigr)
+2 \varepsilon_{\mu \nu}^{\; \; \; \rho \sigma}
\Bigr(\omega_{[\rho} \star {\widetilde \omega}_{\sigma]}
+{\widetilde \omega}_{[\rho} \star \omega_{\sigma]}\Bigr)=0.
\end{eqnarray}

\section{Remaining self-dual equations}

If Eqs. (4.6), (5.3) and (5.4) are fulfilled, we can omit the 
${ }^{+}$ superscript for the curvatures $R_{cd},r,{\widetilde r}$ in
(3.3) and (3.4). Moreover, the form of (3.3) and (3.4) suggests 
defining
\begin{equation}
U_{\mu}^{a} \equiv {\widetilde V}_{\mu}^{a}+V_{\mu}^{a},
\end{equation}
\begin{equation}
W_{\mu}^{a} \equiv {\widetilde V}_{\mu}^{a}-V_{\mu}^{a},
\end{equation}
after which one can use the explicit form of the twist-deformed exterior
product of a $1$-form $\alpha=\alpha_{\lambda}{\rm d}x^{\lambda}$ with a
$2$-form $\gamma={1\over 2}\gamma_{\mu \nu}
{\rm d}x^{\mu} \wedge {\rm d}x^{\nu}$,
i.e.
\begin{equation}
\alpha \wedge_{\star} \gamma={1\over 6}\Bigr(\alpha_{\lambda}
\star \gamma_{\mu \nu}+\alpha_{\mu} \star \gamma_{\nu \lambda}
+\alpha_{\nu} \star \gamma_{\lambda \mu}\Bigr)
{\rm d}x^{\lambda}\wedge {\rm d}x^{\mu} \wedge {\rm d}x^{\nu}.
\end{equation}
Hence one finds that Eq. (3.4) reads as
\begin{equation}
\Bigr[U_{c \lambda} \star ({\widetilde r}_{\mu \nu}+{\rm i}r_{\mu \nu})
+U_{c \mu} \star ({\widetilde r}_{\nu \lambda}+{\rm i}r_{\nu \lambda})
+U_{c \nu} \star ({\widetilde r}_{\lambda \mu}+{\rm i}r_{\lambda \mu})
\Bigr]{\rm d}x^{\lambda}\wedge {\rm d}x^{\mu} \wedge {\rm d}x^{\nu}=0.
\end{equation}
Remarkably, this is also part of Eq. (3.3) in the unknowns 
$U_{\lambda}^{a}$ and $W_{\lambda}^{a}$, which therefore reduces to
an equation in the unknown $W_{\lambda}^{a}$, i.e.
\begin{eqnarray}
\; & \; & \biggr[\Bigr(W_{\lambda}^{d} \star R_{cd \mu \nu}
+W_{\mu}^{d} \star R_{cd \nu \lambda}
+W_{\nu}^{d} \star R_{cd \lambda \mu} \Bigr)
\nonumber \\
&+& \Bigr(W_{c \lambda} \star ({\widetilde r}_{\mu \nu}
-{\rm i}r_{\mu \nu})
+W_{c \mu} \star ({\widetilde r}_{\nu \lambda}
-{\rm i}r_{\nu \lambda})
+W_{c \nu} \star ({\widetilde r}_{\lambda \mu}
-{\rm i}r_{\lambda \mu})\Bigr)\biggr]
{\rm d}x^{\lambda} \wedge {\rm d}x^{\mu} \wedge {\rm d}x^{\nu}
\nonumber \\
&=& 0.
\end{eqnarray}

\section{Self-dual equations to first order in $\theta^{\mu \nu}$}

At this stage we can study the full set of self-dual equations (4.6),
(5.3), (5.4), (6.4) and (6.5) to first order in $\theta^{\rho \sigma}$
as $\theta^{\rho \sigma}$ approaches $0$, since, as we said already
after (1.11), there is no observational evidence so far 
of a regime where noncommutativity produces (even just) 
finite effects, which would justify, in turn,
the consideration of higher orders in $\theta^{\rho \sigma}$. 
Hence we assume existence, in the neighborhood
of $\theta^{\rho \sigma}=0$, of the asymptotic expansion
\begin{equation}
\omega_{\mu}^{ab} \sim { }^{(0)}\omega_{\mu}^{ab}
+\theta^{\rho \sigma}C_{\mu [\rho \sigma]}^{ab}
+{\rm O}(\theta^{2}),
\end{equation}
where ${ }^{(0)}\omega_{\mu}^{ab}$ is the classical spin-connection
from the classical tetrad \cite{DeWi83},
\begin{equation}
{ }^{(0)}\omega_{\mu}^{ab}={1\over 2}e^{a \nu}\Bigr(e_{\nu,\mu}^{b}
-e_{\mu,\nu}^{b}\Bigr)-{1\over 2}e^{b \nu}
\Bigr(e_{\nu,\mu}^{a}-e_{\mu,\nu}^{a}\Bigr)
+{1\over 2}e^{a \nu}e^{b \sigma}\Bigr(e_{\nu,\sigma}^{c}
-e_{\sigma,\nu}^{c}\Bigr)e_{c \mu},
\end{equation}
jointly with (cf. \cite{AC09})
\begin{equation}
\omega_{\mu}(\theta)=-\omega_{\mu}(-\theta) \sim \theta^{\rho \sigma}
A_{\mu [\rho \sigma]}+{\rm O}(\theta^{3}) \; {\rm as} \;
\theta^{\rho \sigma} \rightarrow 0,
\end{equation}
\begin{equation}
{\widetilde \omega}_{\mu}(\theta)=-{\widetilde \omega}_{\mu}(-\theta)
\sim \theta^{\rho \sigma}B_{\mu [\rho \sigma]}
+{\rm O}(\theta^{3}) \; {\rm as} \;
\theta^{\rho \sigma} \rightarrow 0,
\end{equation}
while the asymptotic expansion of $V_{\mu}^{a}$ and 
${\widetilde V}_{\mu}^{a}$ is described by (1.10) and (1.11).

The resulting solution strategy is therefore as follows. First, solve
Eq. (4.6) to first order in $\theta^{\rho \sigma}$, by insertion of
(7.1). Then solve Eqs. (5.3) and (5.4) for $\omega_{\mu}$ and
${\widetilde \omega}_{\nu}$ to first order in $\theta^{\rho \sigma}$,
bearing in mind their limiting form in such a case, i.e.
\begin{equation}
2 \partial_{[\mu} \omega_{\nu]}
+2{\rm i}\varepsilon_{\mu \nu}^{\; \; \; \rho \sigma}
\partial_{[\rho} \omega_{\sigma]}
+{\rm i}\left(-{1\over 4}\omega_{cd [\mu} \star \omega_{\nu ]}^{cd}
-{{\rm i}\over 4}\varepsilon_{\mu \nu}^{\; \; \; \rho \sigma}
\omega_{cd [\rho} \star \omega_{\sigma]}^{cd}\right)
+{\rm O}(\theta^{2})=0,
\end{equation}
\begin{equation}
2 \partial_{[\mu} {\widetilde \omega}_{\nu]}
+2{\rm i}\varepsilon_{\mu \nu}^{\; \; \; \rho \sigma}
\partial_{[\rho} {\widetilde \omega}_{\sigma]}
+\left(-{1\over 4}\omega_{cd [\mu} \star \omega_{\nu ]}^{cd}
-{{\rm i}\over 4}\varepsilon_{\mu \nu}^{\; \; \; \rho \sigma}
\omega_{cd [\rho} \star \omega_{\sigma]}^{cd}\right)
+{\rm O}(\theta^{2})=0.
\end{equation}
At that stage, $R_{\mu \nu}^{ab}$ can be evaluated from (1.22) to 
first order in $\theta^{\rho \sigma}$, as well as 
${\widetilde r}_{\mu \nu} \pm {\rm i}r_{\mu \nu}$ from (1.23) 
and (1.24), i.e.
\begin{equation}
{\widetilde r}_{\mu \nu}+{\rm i}r_{\mu \nu} \sim
2 \Bigr[\partial_{[\mu} {\widetilde \omega}_{\nu ]}
+{\rm i}\partial_{[\mu} \omega_{\nu]}\Bigr]
+{\rm O}(\theta^{2}) \; {\rm as} \; 
\theta^{\rho \sigma} \rightarrow 0,
\end{equation}
\begin{equation}
{\widetilde r}_{\mu \nu}-{\rm i}r_{\mu \nu} \sim
2 \Bigr[\partial_{[\mu} {\widetilde \omega}_{\nu ]}
-{\rm i}\partial_{[\mu} \omega_{\nu]}\Bigr]
-{1\over 2}\omega_{cd [\mu} \star \omega_{\nu]}^{cd}
+{\rm O}(\theta^{2}) \; {\rm as} \;
\theta^{\rho \sigma} \rightarrow 0.
\end{equation}
Thus, one can solve Eq. (6.4) for $U_{\mu}^{a}$, then Eq. (6.5)
for $W_{\mu}^{a}$, and eventually obtain $V_{\mu}^{a}$ and
${\widetilde V}_{\mu}^{a}$ from (6.1) and (6.2), i.e.
$$
V_{\mu}^{a}={1\over 2}(U_{\mu}^{a}-W_{\mu}^{a}), \; 
{\widetilde V}_{\mu}^{a}={1\over 2}(U_{\mu}^{a}+W_{\mu}^{a}).
$$

\section{Solution of Eq. (4.6) to first order}

The insertion of the asymptotic expansion (7.1) into the self-duality
condition (4.6) for the $1$-form $\omega^{ab}$ yields the equation
\begin{equation}
\biggr[\varepsilon_{\; \; \; cd}^{ab}
{ }^{(0)}\omega_{\mu}^{cd}
-2{\rm i}{ }^{(0)}\omega_{\mu}^{ab}\biggr]
=\theta^{\rho \sigma}\biggr[2{\rm i}
C_{\; \; \; \mu [\rho \sigma]}^{ab}
-\varepsilon_{\; \; \; cd}^{ab}
C_{\; \; \; \mu [\rho \sigma]}^{cd}\biggr].
\end{equation}
This condition should be identically satisfied, and we notice that 
the left-hand side is independent of $\theta^{\rho \sigma}$, while the
right-hand side does depend on it. Thus, we should set them to $0$
separately, and a sufficient condition is fulfillment of the
following self-duality conditions (with respect to Lorentz-frame
indices)
\begin{equation}
\varepsilon_{\; \; \; cd}^{ab} 
{ }^{(0)}\omega_{\mu}^{cd}=2{\rm i}
{ }^{(0)}\omega_{\mu}^{ab},
\end{equation}
\begin{equation}
\varepsilon_{\; \; \; cd}^{ab} C_{\mu [\rho \sigma]}^{cd}
=2{\rm i}C_{\mu [\rho \sigma]}^{ab}.
\end{equation}
Interestingly, Eq. (8.2) implies that the curvature $2$-form 
${ }^{(0)}R^{ab}$ of the classical background is itself self-dual. 
Moreover, Eq. (8.2) also implies that,
for a given choice of solution of the self-duality condition for
the classical spin-connection, there exists a $2$-parameter family
of solutions of Eq. (8.3) reading as
\begin{equation}
C_{\mu [\rho \sigma]}^{ab}=\psi_{1} 
{ }^{(0)}\omega_{\mu}^{ab} F_{\rho \sigma}
+\psi_{2}W_{\mu}{ }^{(0)}R_{\rho \sigma}^{ab},
\end{equation}
where $\psi_{1}$ and $\psi_{2}$ are parameters,  
$F_{\rho \sigma}=-F_{\sigma \rho}=F_{[\rho \sigma]}$ are the
components of a generic $2$-form 
$F={1\over 2}F_{\lambda \mu}{\rm d}x^{\lambda} \wedge {\rm d}x^{\mu}$,
and $W_{\mu}$ are the components of a $1$-form $W_{\mu}{\rm d}x^{\mu}$.

\section{Components of the curvature form}

We can begin by studying Eq. (7.6), here re-expressed 
explicitly in the form
\begin{equation}
\partial_{\mu}{\widetilde \omega}_{\nu}
-\partial_{\nu}{\widetilde \omega}_{\mu}
+2{\rm i}\varepsilon_{\mu \nu}^{\; \; \; \rho \sigma}
\partial_{\rho}{\widetilde \omega}_{\sigma} 
- {1\over 8}\biggr(\omega_{cd \mu} \star \omega_{\nu}^{cd}
-\omega_{cd \nu} \star \omega_{\mu}^{cd}
+2 {\rm i}\varepsilon_{\mu \nu}^{\; \; \; \rho \sigma}
\omega_{cd \rho} \star \omega_{\sigma}^{cd} \biggr)
+{\rm O}(\theta^{2})=0.
\label{(9.1)}
\end{equation}
At this stage, we exploit the asymptotic expansion (1.20) of the
twist-deformed product of components of $1$-forms, and the asymptotic
expansion (7.1) of the $1$-form $\omega^{ab}$. Hence we find
the first-order expansion
\begin{eqnarray}
\omega_{cd \mu} \star \omega_{\nu}^{cd} & \sim & 
{ }^{(0)}\omega_{cd \mu} { }^{(0)}\omega_{\nu}^{cd}
+{ }^{(0)}\omega_{cd \mu} \theta^{\alpha \beta} 
C_{\nu [\alpha \beta]}^{cd}
+{ }^{(0)}\omega_{\nu}^{cd} \theta^{\alpha \beta} 
C_{cd \mu [\alpha \beta]}
\nonumber \\
&+& {{\rm i}\over 2}\theta^{\rho \sigma}
\Bigr({ }^{(0)}\omega_{cd \mu,\rho}\Bigr)
\Bigr({ }^{(0)}\omega_{\nu,\sigma}^{cd}\Bigr)
+{\rm O}(\theta^{2}).
\label{(9.2)}
\end{eqnarray}
Interestingly, this first-order asymptotic expansion leads to
exact cancellation of the $4$ terms involving $C_{\mu [\rho \sigma]}^{ab}$
in the course of evaluating, in Eq. (9.1),
the difference among the first $2$ terms within the round bracket
which is multiplied by $-{1\over 8}$. Thus, by virtue of the
asymptotic expansion (7.4), Eq. (9.1) becomes the following
partial differential equation in the unknown $B_{\mu [\rho \sigma]}$:
\begin{eqnarray}
\; & \; & \theta^{\alpha \beta}
\Bigr[B_{\nu [\alpha \beta],\mu}-B_{\mu [\alpha \beta],\nu}
+2{\rm i}\varepsilon_{\mu \nu}^{\; \; \; \rho \sigma}
B_{\sigma [\alpha \beta],\rho}\Bigr]
={{\rm i}\over 8}\theta^{\alpha \beta}  
\Bigr({ }^{(0)}\omega_{cd \mu,\alpha}\Bigr)
\Bigr({ }^{(0)}\omega_{\nu,\beta}^{cd}\Bigr)
\nonumber \\
&+& {{\rm i}\over 4}\varepsilon_{\mu \nu}^{\; \; \; \rho \sigma}
\biggr[{ }^{(0)}\omega_{cd \rho}{ }^{(0)}\omega_{\sigma}^{cd}
+{{\rm i}\over 2}\theta^{\alpha \beta}
\Bigr({ }^{(0)}\omega_{cd \rho,\alpha}\Bigr)
\Bigr({ }^{(0)}\omega_{\sigma,\beta}^{cd}\Bigr)\biggr].
\label{(9.3)}
\end{eqnarray}
Bearing in mind that the Levi-Civita tensor is fully antisymmetric,
and also the identity [see Eq. (10.20)]
$$
{ }^{(0)}\omega_{cd [\rho} { }^{(0)}\omega_{\sigma]}^{cd}=0,
$$
the term independent of $\theta^{\alpha \beta}$ in Eq. (9.3) is found
to vanish, so that this equation reduces to 
\begin{eqnarray}
\; & \; & \theta^{\alpha \beta}
\Bigr[B_{\nu [\alpha \beta],\mu}-B_{\mu [\alpha \beta],\nu}
+2{\rm i}\varepsilon_{\mu \nu}^{\; \; \; \rho \sigma}
B_{\sigma [\alpha \beta],\rho}\Bigr] \nonumber \\
&=& {{\rm i}\over 8}\theta^{\alpha \beta}  
\biggr[\Bigr({ }^{(0)}\omega_{cd \mu,\alpha}\Bigr)
\Bigr({ }^{(0)}\omega_{\nu,\beta}^{cd}\Bigr)
+{\rm i}\varepsilon_{\mu \nu}^{\; \; \; \rho \sigma}
\Bigr({ }^{(0)}\omega_{cd \rho,\alpha}\Bigr)
\Bigr({ }^{(0)}\omega_{\sigma,\beta}^{cd}\Bigr)\biggr]
\nonumber \\
& \equiv & U_{\mu \nu}.
\label{(9.4)}
\end{eqnarray}
Now we define, from the right-hand side of (7.3) and (7.4), 
\begin{equation}
{\cal A}_{\mu} \equiv \theta^{\rho \sigma}A_{\mu [\rho \sigma]}, \;
{\cal B}_{\mu} \equiv \theta^{\rho \sigma}B_{\mu [\rho \sigma]},
\label{(9.5)}
\end{equation}
and we exploit the constancy of $\theta^{\alpha \beta}$ to define
the skew-symmetric `field strengths'
\begin{equation}
G_{\mu \nu} \equiv \partial_{\mu}{\cal A}_{\nu}
-\partial_{\nu}{\cal A}_{\mu}, \;
H_{\mu \nu} \equiv \partial_{\mu}{\cal B}_{\nu}
-\partial_{\nu}{\cal B}_{\mu}.
\label{(9.6)}
\end{equation}
Hence we find, from (7.7) and (7.8), the asymptotic expansions
in the self-dual case
\begin{equation}
{\widetilde r}_{\mu \nu}+{\rm i}r_{\mu \nu}
\sim H_{\mu \nu}+{\rm i}G_{\mu \nu}+{\rm O}(\theta^{2})
\; {\rm as} \; \theta^{\alpha \beta} \rightarrow 0,
\label{(9.7)}
\end{equation}
\begin{equation}
{\widetilde r}_{\mu \nu}-{\rm i}r_{\mu \nu}
\sim H_{\mu \nu}-{\rm i}G_{\mu \nu}
-{{\rm i}\over 4}\theta^{\alpha \beta}
\Bigr({ }^{(0)}\omega_{cd \mu,\alpha}\Bigr)
\Bigr({ }^{(0)}\omega_{\nu,\beta}^{cd}\Bigr)
+{\rm O}(\theta^{2}) \; {\rm as} \;
\theta^{\alpha \beta} \rightarrow 0.
\label{(9.8)}
\end{equation}
In these formulae, $G_{\mu \nu}$ and $H_{\mu \nu}$ are found
by solving equations like (9.4), as is shown in Sec. X. 

Last, but not least, we have to evaluate the curvature components 
$R_{\mu \nu}^{ab}$ from Eq. (1.22) to first order in 
$\theta^{\alpha \beta}$. By virtue of the self-duality assumption
(4.6) and of the asymptotic expansions used so far, we find 
\begin{eqnarray}
R_{\mu \nu}^{ab} & \sim & 2 \biggr[\partial_{[\mu}
{ }^{(0)}\omega_{\nu]}^{ab}
+\theta^{\alpha \beta}
\partial_{[\mu}C_{\nu][\alpha \beta]}^{ab}\biggr]
\nonumber \\
&+& { }^{(0)}\omega_{\; c [\mu}^{b}
{ }^{(0)}\omega_{\nu ]}^{ca}
-{ }^{(0)}\omega_{\; c [\mu}^{a}
{ }^{(0)}\omega_{\nu ]}^{cb}
\nonumber \\
&+&{{\rm i}\over 2}\theta^{\alpha \beta}
\left[\Bigr(\partial_{\alpha}{ }^{(0)}\omega_{\; c [\mu}^{b}\Bigr)
\Bigr(\partial_{\beta} { }^{(0)}\omega_{\nu ]}^{ca}\Bigr)
-\Bigr(\partial_{\alpha} { }^{(0)}\omega_{\; c [\mu}^{a}\Bigr)
\Bigr(\partial_{\beta}{ }^{(0)}\omega_{\nu ]}^{cb}\Bigr)\right]
\nonumber \\
&+& 2 \theta^{\alpha \beta}\biggr[{ }^{(0)}\omega_{[\mu}^{cb}
C_{\; |c|\nu][\alpha \beta]}^{a}
+{ }^{(0)}\omega_{[\nu}^{ca}
C_{\; |c|\mu][\alpha \beta]}^{b}\biggr]
\nonumber \\
&+& 2 \Bigr[{ }^{(0)}\omega_{[\mu}^{ab}
({\cal B}-{\rm i}{\cal A})_{\nu]}
+({\cal B}-{\rm i}{\cal A})_{[\mu}{ }^{(0)}\omega_{\nu]}^{ab}\Bigr]
\nonumber \\
&+& {\rm O}(\theta^{2}) \; {\rm as} \;
\theta^{\alpha \beta} \rightarrow 0,
\label{(9.9)}
\end{eqnarray}
where Eq. (8.4) can be used to express $C_{\mu [\alpha \beta]}^{ab}$.
In the next section we are going to solve Eq. (9.4) and the 
associated equation for ${\cal A}_{\mu}$. This makes it
possible to compute all asymptotic expansions 
of the curvature forms, and the first-order part of the tetrad can
be found eventually from the equation derived in Sec. XI.

\section{Wave equations for ${\cal A}_{\mu}$ and ${\cal B}_{\mu}$} 

Our self-dual solution scheme is fully computable provided that one
is able to obtain the general solution of first-order partial
differential equations like (9.4). For this purpose, we begin by
remarking that Eq. (9.4) can be written in the form 
\begin{equation}
H_{\mu \nu}+2{\rm i}{ }^{(*)}H_{\mu \nu}=U_{\mu \nu},
\label{(10.1)}
\end{equation}
while, from Eq. (7.5), one finds
\begin{equation}
G_{\mu \nu}+2{\rm i}{ }^{(*)}G_{\mu \nu}=-{\rm i} U_{\mu \nu}.
\label{(10.2)}
\end{equation}
Note now that the Hodge dual of Eq. (10.1) yields
\begin{equation}
{ }^{(*)}H_{\mu \nu}-2{\rm i}H_{\mu \nu}={ }^{(*)}U_{\mu \nu},
\label{(10.3)}
\end{equation}
and hence Eqs. (10.1) and (10.3) lead to
\begin{equation}
H_{\mu \nu}
={1\over 3}\Bigr[2{\rm i}{ }^{(*)}U_{\mu \nu}
-U_{\mu \nu}\Bigr].
\label{(10.4)}
\end{equation}
Moreover, since the right-hand side of Eq. (10.2) is $-{\rm i}$
times the right-hand side of Eq. (10.1), we find also
\begin{equation}
G_{\mu \nu}=-{{\rm i}\over 3}\Bigr[2{\rm i}{ }^{(*)}U_{\mu \nu}
-U_{\mu \nu}\Bigr].
\label{(10.5)}
\end{equation}

Note now that, from the point of view of partial differential
equations, Eq. (10.4) can be written explicitly as
\begin{equation}
\partial_{\mu}{\cal B}_{\nu}-\partial_{\nu}{\cal B}_{\mu}
=\nabla_{\mu}{\cal B}_{\nu}-\nabla_{\nu}{\cal B}_{\mu}
={1\over 3}\Bigr[2{\rm i}{ }^{(*)}U_{\mu \nu}
-U_{\mu \nu}\Bigr],
\label{(10.6)}
\end{equation}
where $\nabla_{\mu}$ is a torsion-free metric compatible covariant
derivative of the classical background endowed with classical tetrad
covectors $e_{\; \mu}^{a}$. We need the transition from 
$\partial_{\mu}$ to $\nabla_{\mu}$ because the latter makes it
possible to act with an appropriate derivative operator on both sides
of the tensor equation (10.6), i.e.
\begin{equation}
\nabla^{\mu}\nabla_{\mu}{\cal B}_{\nu}-\nabla^{\mu}\nabla_{\nu}
{\cal B}_{\mu}=-Q_{\nu},
\label{(10.7)}
\end{equation}
having defined
\begin{equation}
Q_{\nu} \equiv -{1\over 3}\nabla^{\mu}\Bigr[2{\rm i}
{ }^{(*)}U_{\mu \nu}-U_{\mu \nu}\Bigr].
\label{(10.8)}
\end{equation}
Eventually, this reads as
\begin{equation}
\Bigr(-\delta_{\nu}^{\mu}\Box+R_{\nu}^{\mu}\Bigr)
{\cal B}_{\mu}+\nabla_{\nu}({\rm div}{\cal B})=Q_{\nu},
\label{(10.9)}
\end{equation}
where $-\delta_{\nu}^{\mu}\Box+R_{\nu}^{\mu}$ is the wave operator
in curved spacetime acting on (co)vectors. It maps elements of
$T_{p}(M)$ into elements of $T_{p}(M)$, and elements of 
$T_{p}^{*}(M)$ into elements of $T_{p}^{*}(M)$; in the language of
differential forms, it reads as ${\rm d}\delta + \delta {\rm d}$,
$\delta$ being the co-differential. Upon imposing the Lorenz 
gauge condition
\begin{equation}
{\rm div}{\cal B}=\nabla^{\mu}{\cal B}_{\mu}=0,
\label{(10.10)}
\end{equation}
Eq. (10.9) becomes the familiar inhomogeneous wave equation
in curved spacetime, for which existence theorems for the solution
are available, since the pioneering work of Leray \cite{Leray} 
on the existence of Green functions of hyperbolic operators 
in curved spacetime \cite{Fried}.

Interestingly, we have therefore found that solutions of the equation 
\begin{equation}
\Bigr(-\delta_{\nu}^{\mu}\Box+R_{\nu}^{\mu}\Bigr)
{\cal B}_{\mu}=Q_{\nu},
\label{(10.11)}
\end{equation}
with ${\cal B}_{\mu}$ satisfying the Lorenz gauge, generate solutions 
of a family of self-dual noncommutative gravity field equations, in
the way made precise by Sec. IX and the following section. 
In particular, on considering classical backgrounds which solve the vacuum
Einstein equations in $4$ dimensions, the Ricci term in Eq. (10.11)
vanishes, and our wave operator takes the simple form
\begin{equation}
P_{\nu}^{\mu} \equiv -\delta_{\nu}^{\mu} \Box,
\label{(10.12)}
\end{equation}
and ${\cal B}_{\mu}$ reads as
\begin{equation}
{\cal B}_{\mu}=b_{\mu}+{\widetilde {\cal B}}_{\mu},
\label{(10.13)}
\end{equation}
where $b_{\mu}$ is the general solution of the homogeneous equation
\begin{equation}
\Box b_{\mu}=0,
\label{(10.14)}
\end{equation}
while ${\widetilde {\cal B}}_{\mu}$ is a particular solution of
the inhomogeneous equation (10.11) with vanishing Ricci term. 
In terms of the Green function $G_{\lambda \nu'} \equiv
G_{\lambda \nu}(x,x')$ of the operator $P_{\mu}^{\lambda}$, which
solves by definition the equation \cite{Bimo04}
\begin{equation}
P_{\mu}^{\lambda}G_{\lambda}^{\; \nu'}=\delta_{\mu}^{\nu}
{\delta(x,x')\over \sqrt{-g}},
\label{(10.15)}
\end{equation}
one finds
\begin{equation}
{\widetilde {\cal B}}_{\mu}=\int G_{\mu}^{\; \nu'} Q_{\nu}(x')
\sqrt{-g(x')}{\rm d}^{4}x'.
\label{(10.16)}
\end{equation}

To obtain an explicit example, we may consider the classical
self-dual spin-connection of a Kasner spacetime \cite{CP}, which
belongs to the class of Bianchi models. In such a case, 
the metric reads as
\begin{equation}
g=-{\rm d}t \otimes {\rm d}t+t^{2p_{1}}{\rm d}x \otimes {\rm d}x
+t^{2p_{2}}{\rm d}y \otimes {\rm d}y 
+t^{2p_{3}}{\rm d}z \otimes {\rm d}z,
\label{(10.17)}
\end{equation}
where the $p_{i}$ are constants satisfying the conditions
\begin{equation}
\sum_{i=1}^{3}p_{i}=1,
\label{(10.18)}
\end{equation}
\begin{equation}
\sum_{i=1}^{3}p_{i}^{2}=1,
\label{(10.19)}
\end{equation}
called the Kasner plane and Kasner $2$-sphere condition, respectively.
Each $t={\rm constant}$ hypersurface of this cosmological model,
which solves the vacuum Einstein equations, is a flat
$3$-dimensional space, and the worldlines of constant $x,y,z$ are
timelike geodesics along which galaxies or other matter, viewed as
test particles, can be imagined to move \cite{MTW73}. This model
represents an expanding universe, since the volume element is 
constantly increasing, but the expansion is anisotropic. The
distances parallel to the $x$-axis expand at a rate proportional
to $t^{p_{1}}$, while those along the $y$-axis can expand at a rate
proportional to $t^{p_{2}}$. Moreover, along one of the axes, 
distances contract rather than expand. Thus, if black-body radiation
were emitted at one time $t$ in a Kasner universe and never 
subsequently scattered, later observers would see blue shifts near
one pair of antipodes on the sky, and red shifts in most other
directions \cite{MTW73}. Despite these features not vindicated by
observations, the model remains of interest, both in
the analysis of classical cosmological singularities 
\cite{Choquet} and for our purposes, 
since we have not a priori reasons for selecting a
particular self-dual solution of the vacuum 
Einstein equations, but we rather
try to build their noncommutative counterpart with the help
of geometric and analytic techniques.

In a Kasner spacetime, the spin-connection satisfies the
self-duality condition (8.2), and its nonvanishing components 
are given by (we use the general formula (7.2), and our
coordinate indices $\mu$ range from $0$ through $3$)
\begin{equation}
{ }^{(0)}\omega_{i}^{ab}=-\Bigr(\delta^{a0}\delta_{i}^{b}-\delta^{b0}
\delta_{i}^{a}\Bigr)p_{i}t^{p_{i}-1}, \;
\forall i=1,2,3.
\label{(10.20)}
\end{equation} 
Thus, the tensor $U_{\mu \nu}$ given by the right-hand side of
Eq. (9.4) is found to vanish (because the term in square brackets
on the second line of (9.4) vanishes if $\alpha \not = \beta$
and $\mu \not = \nu$),
which implies in turn that the field
strengths $H_{\mu \nu}$ and $G_{\mu \nu}$ vanish, by virtue of the
general formulae (10.4) and (10.5). Hence both ${\cal A}_{\mu}$
and ${\cal B}_{\mu}$ {\it can} be expressed as the gradient of one 
and the same scalar function $\phi$, i.e.
\begin{equation}
{\cal A}_{\mu}={\cal B}_{\mu}=\nabla_{\mu}\phi,
\label{(10.21)}
\end{equation} 
and the Lorenz gauge condition upon them leads to the scalar 
wave equation for $\phi$, i.e.
\begin{equation}
\Box \phi={1\over \sqrt{-g}}\partial_{\mu}\Bigr(\sqrt{-g}g^{\mu \nu}
\partial_{\nu}\Bigr)\phi=0.
\label{(10.22)}
\end{equation}
In the Kasner coordinates of Eq. (10.17), this reads as
\begin{equation}
\left[-{\partial^{2}\over \partial t^{2}}
-{1\over t}{\partial \over \partial t}
+t^{-2p_{1}}{\partial^{2}\over \partial x^{2}}
+t^{-2p_{2}}{\partial^{2}\over \partial y^{2}}
+t^{-2p_{3}}{\partial^{2}\over \partial z^{2}}\right]\phi=0.
\label{(10.23)}
\end{equation}
The work in Ref. \cite{Srivastava} suggests looking for solutions
in the form
\begin{equation}
\phi(t,x,y,z)=\int_{-\infty}^{\infty}dk_{1}
\int_{-\infty}^{\infty}dk_{2} \int_{-\infty}^{\infty}dk_{3}
A(k,t) {\rm e}^{{\rm i}(k_{1}x+k_{2}y+k_{3}z)},
\label{(10.24)}
\end{equation}  
where $k$ is a concise notation for the triplet $(k_{1},k_{2},k_{3})$, 
and $A(k,t)$ solves, from (10.23), the partial differential equation
\begin{equation}
\left[-{\partial^{2}\over \partial t^{2}}-{1\over t}
{\partial \over \partial t}-U(k,t)\right]A(k,t)=0,
\label{(10.25)}
\end{equation}
having defined 
\begin{equation}
U(k,t) \equiv \sum_{i=1}^{3}t^{-2p_{i}}k_{i}^{2}.
\label{(10.26)}
\end{equation}
One can turn Eq. (10.25) into a simpler equation, where the coefficient
of the first derivative vanishes, by setting
\begin{equation}
A(k,t)=t^{\alpha}W(k,t).
\label{(10.27)}
\end{equation}
This yields the equation
\begin{equation}
\left[-{\alpha^{2}\over t^{2}}-{(2\alpha+1)\over t}
{{\dot W}\over W}-{{\ddot W}\over W}-U(k,t)\right]A(k,t)=0,
\label{(10.28)}
\end{equation}
where our goal is achieved by setting $\alpha=-{1\over 2}$. Hence
we find that $W$ should solve the equation
\begin{equation}
\left[{\partial^{2}\over \partial t^{2}}+{1\over 4t^{2}}
+U(k,t)\right]W(k,t)=0.
\label{(10.29)}
\end{equation}
To get an understanding of some features of the possible solutions,
we may consider the particular case $p_{1}=1,p_{2}=p_{3}=0$, which is
consistent with the Kasner conditions (10.18) and (10.19). Hence we 
arrive at the equation
\begin{equation}
\left[{\partial^{2}\over \partial t^{2}}+{1\over 4 t^{2}}
+t^{-2}k_{1}^{2}+k_{2}^{2}+k_{3}^{2}\right]W(k,t)=0.
\label{(10.30)}
\end{equation}

The wave equation in Kasner had been studied in Ref. \cite{Hu} 
for a quantum scalar field with mass and conformal coupling term
to gravity, with application to the regularized and renormalized
energy-momentum tensor. Moreover, the classical 
wave equation in a Kasner spacetime had been studied in Ref.
\cite{Goorjan} for the electromagnetic potential, 
where the author obtained plane-wave
solutions such that the temporal component of the electromagnetic
potential vanishes, jointly with $2$ of the spatial components.
The work in Ref. \cite{Sagnotti} had instead evaluated directly  
the electric and magnetic field in Bianchi models, including a 
Kasner universe.

\section{Equations for the tetrad and their solution}

First, by relabelling dummy indices and exploiting the skew-symmetry
of $R_{cd \mu \nu},r_{\mu \nu},{\widetilde r}_{\mu \nu}$ 
and of the exterior product 
${\rm d}x^{\lambda} \wedge {\rm d}x^{\mu}$, we find that the
$3$ terms on the first line of Eq. (6.5) are equal, and the same holds
for the $3$ terms on the second line of Eq. (6.5). Thus, upon defining
\begin{equation}
Z_{ca \mu \nu} \equiv R_{ca \mu \nu}+\eta_{ca}
({\widetilde r}_{\mu \nu}-{\rm i}r_{\mu \nu}),
\label{(11.1)}
\end{equation}
we find that Eq. (6.5) can be expressed in the form
\begin{equation}
W_{\lambda}^{a} \star Z_{ca \mu \nu} {\rm d}x^{\lambda} 
\wedge {\rm d}x^{\mu} \wedge {\rm d}x^{\nu}=0,
\label{(11.2)}
\end{equation}
where, in light of (1.10), (1.11) and (6.2), $W_{\lambda}^{a}$ has
the asymptotic expansion 
\begin{equation}
W_{\lambda}^{a} \sim -e_{\lambda}^{a}
+\theta^{\alpha \beta}P_{\lambda [\alpha \beta]}^{a}
+{\rm O}(\theta^{2}) \; {\rm as} \;
\theta^{\alpha \beta} \rightarrow 0,
\label{(11.3)}
\end{equation}
while, in light of (9.8), (9.9) and (11.1), we write
\begin{equation}
Z_{ca \mu \nu} \sim { }^{(0)}R_{ca \mu \nu}
+\theta^{\alpha \beta} Z_{ca \mu \nu [\alpha \beta]}
+{\rm O}(\theta^{2}) \; {\rm as} \;
\theta^{\alpha \beta} \rightarrow 0,
\label{(11.4)}
\end{equation}
where ${ }^{(0)}R_{ca \mu \nu}$ is the $\theta$-independent
part of the asymptotics (9.9), while
$\theta^{\alpha \beta} Z_{ca \mu \nu [\alpha \beta]}$
is the sum of the parts linear in $\theta$ in the asymptotic
expansions (9.8) and (9.9), i.e.
\begin{eqnarray}
\theta^{\alpha \beta}Z_{ca \mu \nu [\alpha \beta]} & \equiv &
\theta^{\alpha \beta} \biggr \{2 \partial_{[\mu}C_{|ca| \nu][\alpha \beta]}
+{{\rm i}\over 2}
\left[\Bigr(\partial_{\alpha}{ }^{(0)}\omega_{\; ad [\mu}\Bigr)
\Bigr(\partial_{\beta} { }^{(0)}\omega_{|c| \nu ]}^{d}\Bigr)
-\Bigr(\partial_{\alpha} { }^{(0)}\omega_{cd [\mu}\Bigr)
\Bigr(\partial_{\beta}{ }^{(0)}\omega_{|a| \nu ]}^{d}\Bigr)\right]
\nonumber \\
&+& 2 \biggr[{ }^{(0)}\omega_{a [\mu}^{d}
C_{|cd|\nu][\alpha \beta]}
+{ }^{(0)}\omega_{c [\nu}^{d}
C_{|ad|\mu][\alpha \beta]}\biggr]\biggr \}
\nonumber \\
&+& 2 \Bigr[{ }^{(0)}\omega_{ca [\mu}
({\cal B}-{\rm i}{\cal A})_{\nu]}
+({\cal B}-{\rm i}{\cal A})_{[\mu}
{ }^{(0)}\omega_{|ca| \nu]}\Bigr]
\nonumber \\
&+& \eta_{ca}\left[H_{\mu \nu}-{\rm i}G_{\mu \nu}
-{{\rm i}\over 4}\theta^{\alpha \beta}
\Bigr({ }^{(0)}\omega_{pq \mu ,\alpha}\Bigr)
\Bigr({ }^{(0)}\omega_{\nu , \beta}^{pq}\Bigr)\right].
\label{(11.5)}
\end{eqnarray}
Thus, the term $W_{\lambda}^{a} \star Z_{ca \mu \nu}$ 
in Eq. (11.2) is found to have, in the neighborhood of 
$\theta^{\alpha \beta}=0$, the asymptotic expansion
\begin{eqnarray}
\; & \; & W_{\lambda}^{a} \star Z_{ca \mu \nu} \sim
-e_{\lambda}^{a} \; { }^{(0)}R_{ca \mu \nu} \nonumber \\
&+& \theta^{\alpha \beta}\biggr[-e_{\lambda}^{a} 
\; Z_{ca \mu \nu [\alpha \beta]}
+P_{\lambda [\alpha \beta]}^{a}
{ }^{(0)}R_{ca \mu \nu}
-{{\rm i}\over 2} e_{\lambda, \alpha}^{a}
\Bigr({ }^{(0)}R_{ca \mu \nu, \beta}\Bigr)\biggr]
+{\rm O}(\theta^{2}),
\label{(11.6)}
\end{eqnarray}
where the term independent of $\theta^{\alpha \beta}$
on the right-hand side of (11.6) gives a vanishing contribution  
to Eq. (11.2), if the classical background is taken to solve the
vacuum Einstein equations as we have done in Sec. X.
Thus, Eq. (11.2) yields the following `solution' for
$P_{\lambda [\alpha \beta]}^{a}$, which expresses the odd part
of the tetrad in the asymptotic expansion (1.11):
\begin{equation}
P_{\lambda [\alpha \beta]}^{a} \; 
{ }^{(0)}R_{ca \mu \nu}=e_{\lambda}^{a} \; 
Z_{ca \mu \nu [\alpha \beta]}
+{{\rm i}\over 2}e_{\lambda , [\alpha}^{a} \;
{ }^{(0)}R_{|ca \mu \nu|, \beta]}.
\label{(11.7)}
\end{equation}
This equation should be studied jointly with Eq. (6.4), where the 
$3$ terms are equal, so that it reads
\begin{equation}
U_{c \lambda} \star ({\widetilde r}_{\mu \nu}+{\rm i}r_{\mu \nu})
{\rm d}x^{\lambda} \wedge {\rm d}x^{\mu} \wedge {\rm d}x^{\nu}=0.
\label{(11.8)}
\end{equation}
By working to first order in $\theta^{\alpha \beta}$, and introducing
the $2$-forms $G$ and $H$ corresponding to the field strenths
$G_{\mu \nu}$ and $H_{\mu \nu}$, i.e.
\begin{equation}
G \equiv {1\over 2}G_{\mu \nu} {\rm d}x^{\mu} \wedge {\rm d}x^{\nu}, \;
H \equiv {1\over 2}H_{\mu \nu} {\rm d}x^{\mu} \wedge {\rm d}x^{\nu},
\label{(11.9)}
\end{equation} 
Eq. (6.4) leads to the nontrivial restriction
\begin{equation}
e^{c}\wedge (H+{\rm i}G)=0 .
\label{(11.10)}
\end{equation}
As far as we can see, this means that we should choose the solutions
of the wave equations for ${\cal A}_{\mu}$ and ${\cal B}_{\mu}$ in 
such a way that the resulting $2$-forms $G$ and $H$ fulfill Eq. (11.10).
After having checked this, the task remains of solving Eq. (11.7).

In the case of a classical background of the Kasner type, as considered
in the end of Sec. X, both $G$ and $H$ vanish, and hence Eq.
(11.10) is identically satisfied, whereas Eq. (11.7) 
takes a simplified form, obtained by
setting $G_{\mu \nu}=H_{\mu \nu}=0$ and ${\cal A}_{\mu}={\cal B}_{\mu}$
in the formula (11.5). Moreover, in a Kasner background, 
(11.5) is further simplified by the
vanishing of contributions built from partial derivatives of the
classical spin-connection, while (1.22) and (10.20) lead to
the following formulae for nonvanishing components of the classical
curvature $2$-form:
\begin{equation}
{ }^{(0)}R_{0i}^{ab}=-\Bigr(\delta^{a0}\delta_{i}^{b}
-\delta^{b0}\delta_{i}^{a}\Bigr)p_{i}(p_{i}-1)t^{p_{i}-2},
\; \forall i=1,2,3,
\label{(11.11)}
\end{equation}
\begin{equation}
{ }^{(0)}R_{ij}^{ab}=\Bigr(\delta_{i}^{a}\delta_{j}^{b}
-\delta_{j}^{a}\delta_{i}^{b}\Bigr)p_{i}p_{j}
t^{p_{i}+p_{j}-2}, \; \forall i,j=1,2,3.
\label{(11.12)}
\end{equation}
These formulae, bearing also in mind that the tetrad covectors
for the metric (10.17) read as (with no summation over $i$ on the
right-hand-side)
\begin{equation}
e_{\; \lambda}^{a}=\delta_{0}^{a}\delta_{\lambda 0}
+\delta_{i}^{a}t^{p_{i}}\delta_{\lambda i},
\label{(11.13)}
\end{equation}
imply that the skew-symmetrization of partial derivatives on the
right-hand side of Eq. (11.7) vanishes, because the product of such
partial derivatives therein is always proportional to the symmetric
term $\delta_{\alpha 0}\delta_{\beta 0}$. Thus, Eq. (11.7) reduces to
\begin{equation}
P_{\lambda [\alpha \beta]}^{a} \; 
{ }^{(0)}R_{ca \mu \nu}=e_{\lambda}^{a} \; 
Z_{ca \mu \nu [\alpha \beta]},
\label{(11.14)}
\end{equation}
where, on the left-hand side, we read off components of the 
classical curvature $2$-form from (11.11) and (11.12), while on the
right-hand side we use (11.13) for the classical tetrad, and
read off from (11.5) the non-vanishing terms in 
$Z_{ca \mu \nu [\alpha \beta]}$, i.e.
\begin{eqnarray}
\; & \; & Z_{ca \mu \nu [\alpha \beta]} = 
2 \partial_{[\mu}C_{|ca| \nu][\alpha \beta]}
+ 2 \biggr[{ }^{(0)}\omega_{a [\mu}^{d}
C_{|cd|\nu][\alpha \beta]}
+{ }^{(0)}\omega_{c [\nu}^{d}
C_{|ad|\mu][\alpha \beta]}\biggr] \nonumber \\
&+& 2 (1-{\rm i}) \Bigr[{ }^{(0)}\omega_{ca [\mu}
B_{\nu][\alpha \beta]}+B_{[\mu |[\alpha \beta]|}
{ }^{(0)}\omega_{|ca| \nu]}\Bigr].
\label{(11.15)}
\end{eqnarray}
In this expression of $Z_{ca \mu \nu [\alpha \beta]}$ in the
Kasner case, we can set, bearing in mind the definition (9.5),
\begin{equation}
B_{\mu [\alpha \beta]}={\varepsilon_{\alpha \beta}{\cal B}_{\mu}
\over \theta^{\rho \sigma} \varepsilon_{\rho \sigma}},
\label{(11.16)}
\end{equation}
where ${\cal B}_{\mu}$ is obtained from the gradient of the
function (10.24), while the tensor 
$C_{\mu [\rho \sigma]}^{ab}$ admits the general decomposition
displayed in (8.4). We cannot make our solution more explicit
than this. For each choice of $F$ and $W$ in (8.4), we have
a form of $Z_{ca \mu \nu [\alpha \beta]}$, and hence Eqs. 
(11.11)-(11.16) yield algebraic equations for the components 
$P_{\lambda [\alpha \beta]}^{a}$, i.e. the odd part of the tetrad
in the asymptotic expansion (1.11). Our solution task is hence
fully accomplished to first order in $\theta^{\rho \sigma}$.

Note also that, when $\theta^{\rho \sigma}$ has such an orientation
to the $3$ preferred Kasner axes for which only 
$\theta^{yt}$ and $\theta^{zx}$ are nonvanishing
and equal to $\Theta_{1}$ and $\Theta_{2}$, respectively
(the theorem on the reduction to canonical form \cite{Pepe} of
$\theta^{\rho \sigma}$ ensures this is always possible),
its effect reduces to obtaining the following
formula for our $B_{\mu [\alpha \beta]}$:
\begin{equation}
B_{\mu [\alpha \beta]}
={\varepsilon_{\alpha \beta}{\cal B}_{\mu}
\over 2 (\Theta_{1}+\Theta_{2})}.
\end{equation}

\section{Results and open problems} 

In this paper we have tried to develop a powerful `calculus' to find
solutions of the field equations of noncommutative gravity, motivated
by the unsuccessful attempt of applying the Seiberg-Witten map 
\cite{ADE11, DEFV13} to this task when the action functional is built
from twist-deformed exterior products. As far as we know, our 
analysis is original, and its results can be summarized as follows.
\vskip 0.3cm
\noindent
(i) On assuming that the spacetime manifold is parallelizable, so that
tetrads can be introduced, the torsion-free equations resulting from
the action (2.1) take the index-free form (2.7) and (2.8), or, with
Lorentz-frame indices made manifest, the form (2.12) and (2.13).
\vskip 0.3cm
\noindent
(ii) In the self-dual (respectively, anti-self-dual) case, such
equations reduce to (3.3) and (3.4) (respectively, (3.6) and (3.7)).
Self-duality (respectively, anti-self-duality) 
of the $1$-form $\omega^{ab}$ (see (4.6)) is a
sufficient condition for self-duality (respectively, 
anti-self-duality) of the Lorentz-Lie-algebra-valued part of the
full curvature $2$-form. The remaining parts of the curvature
$2$-form are self-dual if the Eqs. (5.3) and (5.4) are satisfied.
The full set of self-dual equations consists of (4.6), (5.3), (5.4),
(6.4) and (6.5).
\vskip 0.3cm
\noindent
(iii) The self-dual equations can be solved by assuming that tetrad 
and connection admit an asymptotic expansion (not of
Poincar\'e type, see appendix D and examples in \cite{Dieu}) 
to first order in noncommutativity in the neighborhood 
of $\theta^{\rho \sigma}$.
This assumption does not exploit the full potentialities of the
twist-deformed exterior product, but might be appropriate after all, 
since no experimental evidence is available as yet of finite
(let alone `large') effects resulting from noncommutativity.
\vskip 0.3cm
\noindent
(iv) Furthermore, all our field
equations can be explicitly solved provided that one is able to
integrate the first-order partial differential equation (9.4),
which turns out to be equivalent to a inhomogeneous wave equation
on $1$-form fields, subject to a Lorenz gauge condition.
\vskip 0.3cm
\noindent
(v) To first order in non-commutativity, the tetrad should fulfill
Eq. (11.7), provided the consistency condition (11.10) is satisfied.
\vskip 0.3cm
\noindent
(vi) The whole scheme has been tested when the classical background
is Kasner spacetime, which is a Bianchi model solving the vacuum 
Einstein equations with self-dual spin-connection. In such a case the
solution of the scalar wave equation (10.23) is the desired
`generator' of a solution for tetrad form and connection form, to
first-order in noncommutativity.

We find it encouraging that the self-dual option can be 
pursued to the extent shown in our paper, without any use of the
Seiberg-Witten map or yet other techniques applied in the previous
literature \cite{2003,2004,2005a,2005b,2005c,2006a,2006b,
2007,2008a,2008b,2008c,2008d,2010,2013}, 
and the nontrivial 
cancellations of terms encountered at some
stages provide further evidence in favour of a new level of internal
consistency of gravity being in sight for the first time. 
Nevertheless, the mathematical potentialities of noncommutative
gravity remain largely unexplored, especially the field equations
and their solutions at finite values of $\theta^{\alpha \beta}$. 

\acknowledgments 
E. Di Grezia and G. Esposito are
grateful to the Dipartimento di Fisica of
Federico II University, Naples, for hospitality and support. 
We are indebted to M. Figliolia for collaboration during the
early stages of this work, to G. Marmo for critical remarks
on the very nature of the geometric objects we have used, and to
P. Aschieri for enlightening correspondence.

\appendix
\section {Twist differential geometry }
\subsection{Twist\label{sectwist}}
\setcounter{equation}{0}
In this section we review the concept of twist, together with 
some of the noncommutative geometry associated to it. 
The presentation is based on Refs. \cite{AC09, A09, ALV08}.  

Let $\Xi$ be the linear space of smooth vector fields on a smooth 
manifold M, and $U\Xi$ its universal enveloping algebra (if $G$ is
a connected Lie group whose Lie algebra ${\cal G}$ is spanned by the
vector fields $\{ L_{\alpha} \}$, the universal enveloping
algebra $U({\cal G})$ is defined to be the algebra generated by the
$L_{\alpha}$'s and the identity, with relations given
by the Lie brackets \cite{Bala}). Given the 
commutative algebra of functions on $M$, denoted by ${\rm Fun}(M)\equiv A$, 
many associative noncommutative products may be obtained from the usual 
pointwise product $\mu(f\otimes g)=fg$ via
the action of a twist operator ${\mathcal F} \in U\Xi\otimes U\Xi$
\begin{equation}
f\star g= \mu\{\mathcal{F}^{-1}(f\otimes g)\}.
\end{equation}
We denote the deformed algebra of functions by $A_\star$.
The associativity of the product is a consequence of the defining 
properties of the twist (an element of  $U\Xi\otimes U\Xi$ is said to 
be a twist if it is invertible, properly normalized and satisfies a 
cocycle condition). On using the standard notation 
\begin{equation}
\mathcal{F}=\mathcal{F}^\alpha\otimes \mathcal{F}_\alpha, 
~~~~~~~ \mathcal{F}^{-1}=\bar{\mathcal{F}}^\alpha
\otimes \bar{\mathcal{F}}_{\alpha},
\end{equation}
with $\mathcal{F}^\alpha,\mathcal{F}_\alpha, \bar{\mathcal{F}}^\alpha, 
\bar{ \mathcal{F}}_\alpha$ elements of $U\Xi$, the star product acquires 
the form
\begin{equation}
f\star g = \bar{\mathcal{F}}^\alpha(f) \bar{\mathcal{F}}_\alpha (g),
\end{equation}
where the elements of $U\Xi$ act on functions as Lie derivatives. 
They are sums of products of vector fields: the Lie derivative with 
respect to products of vector fields is thus extended by means of 
\begin{equation}\mathcal{L}_{v w..}= \mathcal{L}_{v}\mathcal{L}_{ w} 
...\label{iteratedlieder}\end{equation}
 
The class of $\star$-products which can be obtained by a twist is quite 
rich. Among them a wide class is given by the so-called Abelian twists:
\begin{equation}
\mathcal{F}= e^{-\frac{{\rm i}}{2}\theta^{ab} X_a\otimes X_b},
\end{equation}
with $X_a$ mutually commuting vector fields and $\theta^{ab}$ a 
constant antisymmetric matrix. The Moyal twist is a particularly simple 
instance of such a family with $X_a=\partial_a$, the infinitesimal 
generators of translations, globally defined on ${\bf R}^d$.

We also introduce the universal ${\mathcal R}$-matrix
\begin{equation}  {\mathcal R}:={\mathcal F}_{21}{\mathcal F}^{-1},
~\label{defUR} 
\end{equation}
where by definition ${\mathcal F}_{21}:={\mathcal F}_\alpha\otimes 
{\mathcal F}^\alpha$. Hereafter we use
the notation \begin{equation}
{\mathcal R}=R^\alpha\otimes R_\alpha ,~~~~~~
{\mathcal R}^{-1}=\bar{R}^\alpha\otimes\bar{R}_\alpha .
\end{equation}
The ${\mathcal R}$-matrix measures the noncommutativity 
of the $\star$-product. Indeed it is easy to see that 
\begin{equation}\label{Rpermutation} h\star g
=\bar{R}^\alpha(g)\star\bar{R}_\alpha(h).
\end{equation}
The permutation group in noncommutative space is naturally
represented by ${\mathcal R}$. Formula (\ref{Rpermutation}) says that 
the $\star$-product is ${\mathcal R}$-commutative in the sense that, 
if we permute (exchange) two functions by using the ${\mathcal R}$-matrix 
action, then the result does not change.

\subsection{Vector and Tensor fields}

We now use the twist to deform the spacetime commutative geometry
into a noncommutative one. The guiding principle is the
one used to deform the product of functions into the $\star$-product
of functions. Every time we have a bilinear map 
\begin{equation}
\mu\,: X\times Y\rightarrow Z,
\end{equation}
where $X,Y,Z$ are vector spaces, with an action 
of ${\mathcal F}^{-1}$ on $X$ and $Y$, 
we can combine this map with the action of the twist. In
this way we obtain a deformed version, $\mu_\star$,  of the initial
bilinear map $\mu$: 
\begin{eqnarray} \mu_\star:=\mu\circ
{\mathcal F}^{-1},
\label{generalpres}&~~~~~~~~~~~~~~& 
\end{eqnarray} {\vskip -.8cm}
\begin{eqnarray}
{}~~~~~~~~~~~~~\mu_\star\,:X\times  Y&\rightarrow& Z\nonumber\\
(\mathsf{x}, \mathsf{y})\,\, &\mapsto&
\mu_\star(\mathsf{x},\mathsf{y})=\mu(\bar{\mathcal F}^\alpha
(\mathsf{x}),\bar{\mathcal F}_\alpha(\mathsf{y}))\nonumber . 
\end{eqnarray}
The $\star$-product on the space of functions is recovered by setting
$X=Y={A}={\rm Fun}(M)$. We now study the case of vector fields, 
$1$-forms and tensor fields. 
\vskip .5cm
\noindent{\bf Vector fields $\Xi_\star$}.  
\;\;We deform the  $A $-module structure of vector fields, that is the 
product $\mu : A \otimes \Xi\rightarrow \Xi$ between the space
of functions on the spacetime $M$ and vector fields. 
According to the general prescription Eq. (\ref{generalpres}) the product
$\mu : A \otimes \Xi\rightarrow \Xi$ is deformed into the product
\begin{equation}
h\star v=\bar{\mathcal F}^\alpha(h) \bar{\mathcal F}_\alpha(v). 
\end{equation}
The action of $\bar{\mathcal{F}}^\alpha \in U\Xi$ on vector fields is 
given by repeated use of the Lie derivative as in (\ref{iteratedlieder}).
This definition is compatible with the $\star$-product in ${A}$. 
We denote the space of vector fields with this
$\star$-multiplication  by $\Xi_\star$. As vector spaces
$\Xi=\Xi_\star$, but $\Xi$ is an ${ A}$-module while $\Xi_\star$ is an
${\mathcal A}_\star$-module.
\vskip .5cm
\noindent {\bf $1$-forms $\Omega_\star$}. 
\;\;\;Analogously, we deform the
product $\mu : { A}\otimes \Omega\rightarrow \Omega$ between the
space ${A}$ of functions on spacetime $M$ and
$1$-forms. As for vector fields, we have
\begin{equation}
h\star \rho=\bar{\mathcal F}^\alpha(h) \bar{\mathcal F}_\alpha(\rho).
\end{equation}
The action of $\bar{\mathcal F}_\alpha$ on forms is given by iterating the
Lie derivative action of vector fields on forms, as a trivial 
generalization of Eq. (\ref{iteratedlieder}). Forms can be multiplied 
by functions from the left or from the right (they are an $A$ bimodule). 
If we deform the multiplication from the right we obtain the new product
\begin{equation}
 \rho \star h =\bar{\mathcal F}^\alpha(\rho) 
\bar{\mathcal F}_\alpha(h) ,
\end{equation}
and we move $h$ to the left with the help of the ${\mathcal R}$-matrix,
\begin{equation}
 \rho \star h = \bar R^\alpha(h)\star \bar R_\alpha(\rho). 
\end{equation}
We have therefore defined the $A_\star$-bimodule of $1$-forms.
\vskip .5cm

\noindent {\bf Tensor fields {${\mathcal T}_\star$}}.
\;\; Tensor fields form an
algebra with the tensor product $\otimes$ (over the algebra of
functions). We define ${\mathcal T}_\star$ to be the noncommutative
algebra of tensor fields. As vector spaces 
${\mathcal T}={\mathcal T}_\star$; the
noncommutative and associative tensor product is obtained by
applying (\ref{generalpres}):  
\begin{equation}
\label{defofthetensprodst}
\tau\otimes_\star\tau':=\bar{\mathcal F}^\alpha(\tau)\otimes
\bar{\mathcal F}_\alpha(\tau') . 
\end{equation}
Here again the action of the twist on
tensors is via the Lie derivative. Use of the Leibniz rule gives 
the action of the Lie derivative on a generic tensor.

There is a natural action of the permutation group on undeformed
arbitrary tensor fields:
\begin{equation}
\tau\otimes\tau' \stackrel{\sigma}{\longrightarrow}
\tau'\otimes\tau .
\end{equation}
In the deformed case it is the ${\mathcal R}$-matrix that provides a
representation of the permutation group on $\star$-tensor fields:
\begin{equation}
\tau\otimes_\star\tau'\stackrel{\sigma_{_{\mathcal R}}}{\longrightarrow}
\bar{R}^\alpha(\tau')\otimes_\star \bar{R}_\alpha(\tau) .
\end{equation}
It is easy to check that, consistently with $\sigma_{\mathcal R}$ being a
representation of the permutation group, we have
$(\sigma_{\mathcal R})^{2}={\rm id}$.

\vskip .5cm
\noindent{\bf Exterior forms $\Omega_\star^{\circ}
=\oplus_p\Omega_\star^p$}. Exterior forms form and algebra with product 
$\wedge: \Omega^\circ\times\Omega^\circ\rightarrow \Omega^\circ$. According 
to the general prescription (\ref{generalpres}) we $\star$-deform the 
wedge product
\begin{equation}
\theta \wedge_\star\theta'=\bar{\mathcal F}^\alpha(\theta)
\wedge \bar{\mathcal F}_\alpha(\theta').
\end{equation}
As a particular instance of the tensor product above, the exterior product 
is associative and $\bar {\mathcal F}^\alpha, \bar {\mathcal F}_\alpha$ act 
as Lie derivatives. Therefore, the exterior derivative ${\rm d}$, 
commuting with the Lie derivative, is undeformed and satisfies the standard 
graded Leibniz rule
\begin{equation}
{\rm d} (\theta\wedge_\star \theta')={\rm d}\theta \wedge_\star \theta'
+(-1)^{{\rm deg}(\theta)} \theta\wedge_ \star {\rm d} \theta'.
\end{equation}

For Abelian twists constructed with globally defined vector fields 
the ordinary integral of forms verifies the graded cyclicity property, 
that is, up to boundary terms,
\begin{equation}
\int \theta\wedge_\star \theta'= \int \theta \wedge \theta'
= (-1)^{ \deg (\theta) \deg (\theta')}
\int  \theta' \wedge \theta= (-1)^{\deg (\theta) 
\deg (\theta')}\int  \theta' \wedge_\star \theta ,
\end{equation}
with $\theta\wedge \theta'$ a form of maximal rank on the spacetime 
manifold. It is possible to show that the graded cyclicity holds for more 
general twists. 

As for complex conjugation, we have, for Abelian twists defined in terms 
of real fields $X_a$, 
\begin{equation}
(\theta\wedge_\star \theta')^* = (-1)^{\deg(\theta) 
\deg(\theta')} \theta'^* \wedge_\star \theta^{*},
\end{equation}
which holds in particular for functions. 

\subsection{Infinitesimal $\star$-diffeomorphisms}

We have mentioned  in Sec. \ref{section2} that the gravity action 
Eq. (\ref{action}) is invariant under standard diffeomorphisms, which 
are generated by vector fields, 
that act on forms through the Lie derivative. Indeed we have
\begin{equation}
\mathcal{L}_v\int 4-{\rm form}= 
\int {\rm d}( i_{v} 4-{\rm form}),  \label{boundary}
\end{equation}
which yields a boundary term. Interestingly, the $\star$-action 
in (\ref{action}) is also invariant with respect to $\star$-diffeomorphisms.  
Let us describe the $\star$-Lie algebra structure of their infinitesimal 
generators. 

Following the general prescription (\ref{generalpres}) we may  
combine the usual Lie derivative on functions ${\mathcal
L}_uh=u(h)$ with the twist ${\mathcal F}$ 
\begin{equation}
\label{stliederact} 
{\mathcal L}^\star_u(h):=\bar{\mathcal F}^\alpha(u)
(\bar{\mathcal F}_\alpha(h)).
\end{equation}
We obtain in this way the $\star$-Lie derivative on the algebra of 
functions $A _\star$. The differential operator 
${\mathcal L}^\star_u$ satisfies the deformed Leibniz rule
\begin{equation}
{\mathcal L}^\star_u(h\star g)={\mathcal
L}_u^\star(h)\star g + \bar{R}^\alpha(h)\star {\mathcal
L}_{\bar{R}_\alpha(u)}^\star(g) . 
\label{defL}
\end{equation}
This deformed Leibniz rule is intuitive: in the second addend
we have exchanged the order of $u$ and $h$, and this is
achieved by the action of the ${\mathcal R}$-matrix, which
provides a representation of the permutation group. In the commutative 
case the commutator of two vector fields
is again a vector field, we have the Lie algebra of vector fields. 
In this $\star$-deformed case we have a similar situation.
It is possible to verify that 
\begin{equation}
\label{stLieastrep} 
{\mathcal L}^\star_u\,{\mathcal L}^\star_v-
{\mathcal L}^\star_{\bar R^\alpha (v)}\,{\mathcal L}^\star_{\bar
R_\alpha(u)}={\mathcal L}^\star_{[u,v]_\star},
\end{equation}
where we have defined the new vector field 
\begin{equation}
[u,v]_\star := [ \bar{\mathcal F}^\alpha(u) , 
\bar{\mathcal F}_\alpha(v)], 
\label{etwa}
\end{equation}
again as in (\ref{generalpres})
the deformed bracket is obtained from
the undeformed one via composition with the twist: 
\begin{equation}
[~,~]_\star=[~,~] \circ{\mathcal F}^{-1}.
\label{etwa2}
\end{equation}
Therefore, in the presence of twisted noncommutativity, we associate to the 
usual Lie algebra of vector fields, 
$\Xi$, $\Xi_\star$, the algebra of vector fields 
equipped with the $\star$-bracket (\ref{etwa}) or equivalently (\ref{etwa2}).  
The map 
$
[~~,~~]_\star~:~\Xi_\star \times\Xi_\star \rightarrow \Xi_\star\nonumber
$
is a bilinear map and verifies the $\star$-antisymmetry and the
$\star$-Jacoby identity
\begin{equation}\label{sigmaantysymme}
[u,v]_\star =-[\bar{R}^\alpha(v), \bar{R}_\alpha(u)]_\star ,
\end{equation}
\begin{equation}\label{stJac}
[u,[v,z]_\star ]_\star =[[u,v]_\star ,z]_\star
+ [\bar{R}^\alpha(v), [\bar{R}_\alpha(u),z]_\star ]_\star .
\end{equation}
We have constructed the deformed Lie algebra of vector fields
$\Xi_\star$. As vector spaces $\Xi=\Xi_\star$, but $\Xi_\star$ is a 
$\star$-Lie algebra. We stress that a $\star$-Lie algebra is not a 
generic name for a deformation of a Lie algebra. 
Rather it is a quantum Lie algebra of a quantum (symmetry) 
group \cite{Woronowicz}. 

Eq. (\ref{defL}) makes  vector fields into  $\star$-derivations of 
$\mathcal{A}_\star$. Moreover, it is compatible with the 
$\star$-multiplication on the left by elements of $\mathcal{A}_\star$, 
making  $\Xi_\star$ into a  left ${\mathcal A}_\star$- module.

$\star$-vector fields are the infinitesimal generators of 
$\star$-diffeomeorphisms. It is not difficult to verify that the action 
(\ref{action}) is  invariant. We have indeed
\begin{equation}
\mathcal{L}^\star_v\int {\rm 4-form}= 
\int \mathcal{L}^\star_v ({\rm 4-form})= 
\int \mathcal{L}_{\bar{\mathcal{F}^\alpha}(v) }  
\bar{\mathcal{F}_\alpha} (\rm{4-form}).
\end{equation}
On using Eq. (\ref{iteratedlieder})  to compute 
$\mathcal{L}_{\bar{\mathcal{F}^\alpha}(v) }$ and observing that 
$\bar{\mathcal{F}_\alpha}$ itself acts on forms as a Lie derivative, 
we end up with the integral of the external derivative of a top form, 
as in Eq. (\ref{boundary}), 
which yields again a  boundary term.

\section{$\star$-gauge transformations and traces in the field equations}

The form of the expansions (1.8), (1.12) and (1.16) can be
understood by making the following considerations \cite{AC09}.
If two infinitesimal gauge transformations $\tau$ and $\tau'$
are given, reading as 
\begin{equation}
\tau=I+\delta \varepsilon, \; \tau'=I+\delta \varepsilon',
\end{equation}
where $\varepsilon=\varepsilon^{A}T^{A}$ and $T^{A}$ are the generators
of the algebra of the group under consideration, 
the deformed commutator of $\tau$ and $\tau'$ can be 
expressed in the form
\begin{equation}
[\tau,\tau']_{\star}=[\delta \varepsilon,\delta \varepsilon']_{\star}
={1\over 2} \left \{ \delta \varepsilon^{A},\delta \varepsilon'^{B}
\right \}_{\star}[T^{A},T^{B}]
+{1\over 2}[\delta \varepsilon^{A},\delta \varepsilon'^{B}]_{\star}
\left \{T^{A},T^{B} \right \}.
\end{equation}
Thus, since the deformed commutator of infinitesimal gauge
parameters does not vanish, for a generic Lie algebra it is
necessary to perform an extension so as to include also the
anti-commutators of generators, hence considering all their 
possible products. 

In the specific case of the spinor representation of the Lorentz
group, the expansion of the noncommutative gauge parameter
$\varepsilon$ contains also contributions proportional to the identity
matrix $I$ and to $\gamma_{5}$, i.e.
\begin{equation}
\varepsilon=\epsilon^{ab}\Gamma_{ab}+{\rm i}{\hat \epsilon}I
+{\tilde \epsilon}\gamma_{5},
\end{equation}
where the new parameters ${\hat \epsilon}$ and ${\tilde \epsilon}$,
absent in the commutative setting, can be chosen to be real as the
remaining ones, which is equivalent to imposing the 
Hermiticity condition
\begin{equation}
-\gamma_{0}\varepsilon \gamma_{0}={\varepsilon}^{\dagger}.
\end{equation}
Thus, to achieve closure of noncommutative gauge transformations,
the original Lorentz group $SO(3,1)$ of commutative theory has
been extended to the group 
$SO(3,1) \times U(1) \times {\bf R}^{+}$, where the matrices 
${\rm i}I$ and $\gamma_{5}$ are the generators of the compact
component $U(1)$ and noncompact component ${\bf R}^{+}$,
respectively. More precisely, the original gauge group $SL(2,C)$
has been therefore extended to the $\star$-gauge group $GL(2,C)$.

Since, under infinitesimal $\star$-gauge transformations, the full
connection form $\Omega$ (called spin-connection) undergoes the
variation
\begin{equation}
\delta_{\varepsilon}\Omega=\delta \varepsilon
-[\Omega,\varepsilon]_{\star}, n\label{delOmega}
\end{equation}
it takes values, jointly with the curvature $2$-form, in the
$GL(2,C)$ Lie algebra given by even products of $\gamma$-matrices,
according to the expansions (1.12) and (1.16), respectively.
The reality conditions for the $1$-forms $\omega$, 
${\widetilde \omega}$, and the $2$-forms $r$, ${\widetilde r}$, 
can be summarized through the Hermiticity conditions
\begin{equation}
-\gamma_{0}\Omega \gamma_{0}=\Omega^{\dagger}, \;
-\gamma_{0}{\cal R}\gamma_{0}={\cal R}^{\dagger}.
\end{equation}

Moreover, the infinitesimal $\star$-gauge transformations for
tetrads read as
\begin{equation}
\delta_{\varepsilon}V=-[V,\varepsilon]_{\star}, \label{delV}
\end{equation}
and they `close' in the linear space generated by odd 
$\gamma$-matrices, i.e. both $\gamma_{a}$ and 
$\gamma_{a}\gamma_{5}$, the latter resulting from the 
anti-commutator $\left \{ \gamma_{ab}, \gamma_{c} \right \}$.
Hence one arrives at the expansion (1.8). 

On using Eqs. (\ref{delOmega}), (\ref{delV}) it can be easily verified 
that the variation of the action (\ref{action}) with respect to 
$\star$-gauge transformations vanishes \cite{AC09} that is, the model 
is $\star$-gauge invariant, with gauge group $GL(2,C)$. 

We also find it helpful for the general reader to 
evaluate the $6$ traces
which contribute to the field equation (2.10), i.e. \cite{AC09}
\begin{equation}
\tau_{1} \equiv {\rm Tr}\biggr \{ {{\rm i}\over 4}
\gamma_{c}\gamma_{5}\Bigr(V^{d}\gamma_{d} \wedge_{\star} R^{ab}
\gamma_{ab}\Bigr)\biggr \}
=-{{\rm i}\over 4}{\rm Tr}(\gamma_{ab}\gamma_{c}\gamma_{d}\gamma_{5})
V^{d} \wedge_{\star}R^{ab}
=-\varepsilon_{abcd}V^{d}\wedge_{\star}R^{ab},
\end{equation}
\begin{equation}
\tau_{2} \equiv {\rm Tr} \biggr \{-\gamma_{c}\gamma_{5}V^{d}
\gamma_{d} \wedge_{\star} r \biggr \}
={\rm Tr}(\gamma_{c}\gamma_{d}\gamma_{5})V^{d} \wedge_{\star}r=0,
\end{equation}
\begin{equation}
\tau_{3} \equiv {\rm Tr} \biggr \{ {\rm i}\gamma_{c}\gamma_{5}
V^{d}\gamma_{d} \wedge_{\star} {\widetilde r} \gamma_{5} \biggr \}
=-{\rm i} {\rm Tr}(\gamma_{c}\gamma_{d})V^{d} \wedge_{\star}
{\widetilde r}=-4{\rm i}V_{c} \wedge_{\star} {\widetilde r},
\end{equation}
\begin{equation}
\tau_{4}\equiv {\rm Tr} \biggr \{ {{\rm i}\over 4}\gamma_{c}\gamma_{5}
{\widetilde V}^{d}\gamma_{d}\gamma_{5} \wedge_{\star} R^{ab}
\gamma_{ab} \biggr \}
=-{{\rm i}\over 4}{\rm Tr}(\gamma_{c}\gamma_{d}\gamma_{ab})
{\widetilde V}^{d} \wedge_{\star} R^{ab}
=-{\rm i}(\eta_{bc}\eta_{ad}-\eta_{ac}\eta_{bd})
{\widetilde V}^{d} \wedge_{\star} R^{ab},
\end{equation}
\begin{equation}
\tau_{5} \equiv {\rm Tr} \biggr \{
-\gamma_{c} \gamma_{5} {\widetilde V}^{d}\gamma_{d}\gamma_{5}
\wedge_{\star} r \biggr \}
={\rm Tr}(\gamma_{c}\gamma_{d}){\widetilde V}^{d} \wedge_{\star} r
=4 {\widetilde V}_{c} \wedge_{\star}r,
\end{equation}
\begin{equation}
\tau_{6} \equiv \biggr \{ {\rm i}\gamma_{c}\gamma_{5}
{\widetilde V}^{d}\gamma_{d}\gamma_{5} \wedge_{\star}
{\widetilde r} \gamma_{5} \biggr \}=0.
\end{equation}

\section{The Hodge dual}

Our definitions of Hodge duals (1.17) and (1.18) are inspired by
earlier work in the literature. For example, the work in Ref.
\cite{Eguchi} used precisely the definition (1.17) to derive
self-dual solutions of Euclidean gravity, i.e. the asymptotically
locally Euclidean Eguchi--Hanson instanton. What is nontrivial in 
these definitions is that $\omega^{ab}$ is a $1$-form
$\omega_{\mu}^{ab}{\rm d}x^{\mu}$ but, being Lie-algebra-valued
and skew-symmetric: $\omega^{ab}=-\omega^{ba}$, makes it possible
to define a Hodge dual (1.17) with respect to Lorentz-frame indices. 
The same holds for $R^{ab}$ which is a $2$-form written after 
Eq. (1.16). Our Levi-Civita symbol with frame indices,
$\varepsilon_{abcd}$, is precisely the one used in Ref. 
\cite{AC09}, i.e. the standard undeformed Levi-Civita symbol
with frame indices, obtainable from flat-space $\gamma$-matrices
according to
\begin{equation}
\varepsilon_{abcd}={{\rm i}\over 4}{\rm Tr}
(\gamma_{ab}\gamma_{c}\gamma_{d}\gamma_{5}).
\end{equation}

By contrast, the $2$-forms $r$ and ${\widetilde r}$ introduced in
Sec. I are Lorentz scalars, or, in other words, $0$-forms from the point
of view of frame indices, and hence for them we have to generalize the
definition of Hodge dual in curved spacetime. Indeed, in Riemannian
geometry, the Hodge dual of a $2$-form $\alpha$ ($\beta$ being another
$2$-form) admits the intrinsic definition
\begin{equation}
{ }^{(*)}\alpha \wedge \beta=(\alpha,\beta)V_{4},
\end{equation}
where $(\alpha,\beta)$ is the interior product of $\alpha$ with
$\beta$, and $V_{4}$ is the volume $4$-form. With index notation,
one then writes \cite{GegDas}
\begin{equation}
({ }^{(*)}\alpha)_{\mu \nu}={1\over 2}
\varepsilon_{\mu \nu}^{\; \; \; \rho \sigma}\alpha_{\rho \sigma},
\end{equation}
where $\varepsilon_{\mu \nu \rho \sigma} \equiv 
\sqrt{{\rm det} \; g} \; \epsilon_{\mu \nu \rho \sigma}$, where
$\epsilon_{\mu \nu \rho \sigma}$ is equal to $1$ (resp. $-1$)
for even (resp. odd) permutation of the indices, and equal
to $0$ otherwise. The `curved' Levi-Civita symbol 
$\varepsilon_{\mu \nu \rho \sigma}$ is a covariant tensor density
of weight $-1$, whereas $\varepsilon^{\mu \nu \rho \sigma}$ is
a contravariant tensor density of weight $+1$. Last, but not least,
the Levi-Civita symbol with a pair of covariant and a pair of
contravariant indices is a tensor of type $(2,2)$, skew-symmetric
in both pairs of indices.

Both definitions recalled so far make it quite clear that, to
define the Hodge dual, one needs a metric. In Lorentzian geometry, 
the metric has signature $2$ in dimension $4$, and hence the Hodge
dual becomes a complex structure, its square being equal to minus
the identity.

Within the framework of twist differential geometry applied to
gravity, we know that the tensor product gets deformed according to 
our prescription (A15). Thus, we deform the tensor product 
$e^{a} \otimes e^{b}$ in Eq. (1.3), after pointing out that
the expansion (1.8) can be written in the form 
\begin{equation}
V_{\mu}=E_{\; \mu}^{a} \gamma_{a},
\end{equation}
where
\begin{equation}
E_{\; \mu}^{a} \equiv V_{\; \mu}^{a}I
-{\widetilde V}_{\; \mu}^{a} \gamma_{5},
\end{equation}
or, with matrix indices made explicit,
\begin{equation}
(E_{\; \mu}^{a})_{j}^{\; l} \equiv 
V_{\; \mu}^{a} \delta_{j}^{\; l}
-{\widetilde V}_{\; \mu}^{a}(\gamma_{5})_{j}^{\; l}.
\end{equation}
Thus, what corresponds to the tetrad $1$-forms $e^{a}$ of
Eq. (1.2) is the matrix of $1$-forms 
$E^{a}=E_{\; \mu}^{a}{\rm d}x^{\mu}$, and the previous 
considerations suggest considering the following definition of
metric (the factor ${1\over 4}$ is introduced to compensate
for ${\rm Tr}I=4$):
\begin{equation}
g \equiv {1\over 4}{\rm Tr}(E^{a} \otimes_{\star}E^{b})\eta_{ab}
=g_{\mu \nu}(\theta){\rm d}x^{\mu} \otimes {\rm d}x^{\nu},
\end{equation}
where
\begin{equation}
g_{\mu \nu}(\theta)={1\over 4}{\rm Tr}
(E_{\; \mu}^{a} \star E_{\nu}^{b}) \eta_{ab}
=\Bigr(V_{\; \mu}^{a} \star V_{\; \nu}^{b}
+{\widetilde V}_{\; \mu}^{a} \star
{\widetilde V}_{\; \nu}^{b}\Bigr)\eta_{ab},
\end{equation}
which reduces to (1.1) as $\theta \rightarrow 0$.
Furthermore, we note that, similarly to 
the way in which the undeformed tetrad $e_{\; \mu}^{a}$ turns
the Levi-Civita symbol (C1) into its curved 
spacetime counterpart according to
\begin{equation}
\epsilon_{\mu \nu \rho \sigma} =
\varepsilon_{abcd} \; e_{\; \mu}^{a} \; e_{\; \nu}^{b}
\; e_{\; \rho}^{c} \; e_{\; \sigma}^{d},
\end{equation}
we can now define, with the help of $E_{\; \mu}^{a}$,
\begin{eqnarray}
\; & \; &
E_{\mu \nu \rho \sigma} \equiv {\rm Tr}
\Bigr[\varepsilon_{abcd}(E_{\; \mu}^{a} \star E_{\; \nu}^{b}
\star E_{\; \rho}^{c} \star E_{\; \sigma}^{d})\Bigr]
\nonumber \\
&=& \varepsilon_{abcd}(E_{\; \mu}^{a})_{j}^{\; l}
\star (E_{\; \nu}^{b})_{l}^{\; m}
\star (E_{\; \rho}^{c})_{m}^{\; p}
\star (E_{\; \sigma}^{d})_{p}^{\; j}.
\end{eqnarray}
To raise and lower indices of $E_{\mu \nu \rho \sigma}$, some
equally legitimate (but different) prescriptions are available, i.e.
\begin{equation}
E_{\mu \nu}^{\; \; \; \rho \sigma} \equiv
(g^{\rho \alpha} \star g^{\sigma \beta} \star
E_{\mu \nu \alpha \beta}) \; {\rm or} \;
(g^{\rho \alpha} \star E_{\mu \nu \alpha \beta} \star
g^{\beta \sigma}) \; {\rm or} \;
(E_{\mu \nu \alpha \beta} \star g^{\alpha \rho} \star
g^{\beta \sigma}),
\end{equation}
as well as other prescriptions differing for the relative order
of indices of metric components. Some freedom is also available
in the definition of contravariant metric $g^{\mu \nu}$, i.e.
\begin{equation}
g_{\mu \nu} \star g^{\nu \lambda}=\delta_{\mu}^{\; \lambda} 
\; {\rm or} \; 
g^{\mu \nu} \star g_{\nu \lambda}=\delta_{\; \lambda}^{\mu}.
\end{equation}
Last, to define the Hodge dual of a $2$-form when curved spacetime
is deformed according to the prescriptions of twist differential
geometry, we consider (cf. Ref. \cite{GegDas}) 
\begin{equation}
{\cal E}_{\mu \nu}^{\; \; \; \rho \sigma} \equiv 
\sqrt{{\rm det} \; g(\theta)} \star
E_{\mu \nu}^{\; \; \; \rho \sigma},
\end{equation}
where we define, inspired by matrix calculus,
\begin{equation}
{\rm det} \; g(\theta) \equiv E^{i_{1}...i_{4}}
\star g_{1i_{1}} \star ... \star g_{4 i_{4}},
\end{equation}
with the understanding that $E^{i_{1}...i_{4}}$ is defined
according to (C10), and the metric components are defined
according to (C8). We propose therefore the following definition
of Hodge dual of a $2$-form $\alpha={1\over 2}\alpha_{\mu \nu}
{\rm d}x^{\mu} \wedge {\rm d}x^{\nu}$:
\begin{equation}
({ }^{(*)}\alpha)_{\mu \nu} \equiv 
{1\over 2}{\cal E}_{\mu \nu}^{\; \; \; \rho \sigma}
\star \alpha_{\rho \sigma} \sim
{1\over 2}\varepsilon_{\mu \nu}^{\; \; \; \rho \sigma}
\alpha_{\rho \sigma}
+({\rm O}(\theta)\alpha)_{\mu \nu},
\end{equation}
where $\varepsilon_{\mu \nu}^{\; \; \; \rho \sigma}$ includes
$\sqrt{{\rm det} \; g(\theta=0)}$. 

The transformation properties of our geometric objects are defined
when one considers their behavior under infinitesimal
$\star$-diffeomorphisms studied in section 3 of our appendix A.
This involves studying the $\star$-Lie derivative of deformed
products according to our Eq. (A24). In our application to
gravity, we shall therefore write Eq. (A24) in the form
\begin{equation}
{\rm Tr}\Bigr[{\cal L}_{u}^{\star}(E_{\; \mu}^{a}
\star E_{\; \nu}^{b})\Bigr]
={\rm Tr}\Bigr[({\cal L}_{u}^{\star}E_{\; \mu}^{a})
\star E_{\; \nu}^{b}
+{\overline R}^{\alpha}(E_{\; \mu}^{a}) \star
{\cal L}_{{\overline R}_{\alpha}(u)}^{\star}
E_{\; \nu}^{b}\Bigr].
\end{equation}
The passage to some sort of `exponentiation' to obtain the full set
of finite $\star$-diffeomorphisms is a challenging open problem,
as far as we know.

Interestingly, our definition expressed by (C15) leads
to our asymptotic expansions (3.2), which tell us that, to first
order in $\theta$, since both $r_{\mu \nu}$ and
${\widetilde r}_{\mu \nu}$ are odd functions of $\theta$, one can 
keep using the Levi-Civita symbol with coordinate indices of
the undeformed curved spacetime. In general, {\it for the purpose
of studying linear effects of} $\theta$, {\it the various conceivable 
definitions of Hodge dual of a} $2$-{\it form lead always to the 
asymptotic expansions (3.2)}.
   
\section{Asymptotic expansions}

Following \cite{Dieu}, we find it appropriate to stress that the
notion of asymptotic expansion has nothing to do with the notion of
series, despite the confusing use of the term `asymptotic series' in
the literature. A series has infinitely many terms, whereas, by
definition (see below), {\it an asymptotic expansion has only finitely
many terms}. Talking about convergence (or lack of) 
of an asymptotic expansion is therefore meaningless. The confusion
arises because, in several cases, the Taylor expansion in the
neighborhood of a real point $x_{0}$ of the function under 
consideration can be extended arbitrarily far away from $x_{0}$,
and one can then try to understand whether the Taylor series
converges and what is the relation between its sum and the function
one started from. This problem, however, has no relation whatsoever
with the study of the behavior of the given function in the 
neighborhood of $x_{0}$.
 
The existence of asymptotic expansions with 
a large number of terms is a very
special phenomenon. For example, the function
$$
x \rightarrow x^{2}+x \sin x 
$$
has an asymptotic expansion with one term only, i.e.
$x^{2}+o(x^{2})$, in the neighborhood of $+\infty$. Another
example is provided by the number $\pi(x)$ of prime numbers
smaller than or equal to $x$, for which 
$$
\pi(x) \sim \int_{2}^{x}{{\rm d}t \over \log t}.
$$

In general, one starts by considering the set ${\cal E}$ of functions
of the form \cite{Dieu}
\begin{equation}
g: x \rightarrow g(x) \equiv x^{\alpha}(\log x)^{\beta}
{\rm e}^{P(x)},
\end{equation}
where $\alpha,\beta$ are real nonvanishing constants, and
\begin{equation}
P(x) \equiv \sum_{j=1}^{k} c_{j}x^{\gamma_{j}},
\end{equation}
where the $c_{j}$ are real constants of arbitrary sign, while
\begin{equation}
\gamma_{1} > \gamma_{2} > ... > \gamma_{k}>0.
\end{equation}

By definition, given a function $f$, its asymptotic expansion with
$k$ terms with respect to the set ${\cal E}$ is meant to be the sum
\cite{Dieu} 
\begin{equation}
\Sigma_{k} \equiv \sum_{j=1}^{k}b_{j}g_{j} ,
\end{equation}
where the $b_{j}$ are nonvanishing constants, and 
the $g_{j}$ are functions belonging to
the set ${\cal E}$ such that
\begin{equation}
g_{j+1}=o(g_{j}), \; \forall j: 1 \leq j \leq k-1.
\end{equation}
One then writes
\begin{equation}
f= \sum_{j=1}^{k}b_{j}g_{j}+o(g_{k}).
\end{equation}
The difference $f-\Sigma_{k}$ is called the {\it remainder}
of the asymptotic expansion \cite{Dieu}. In our paper we write
this last formula with the equality symbol replaced by the
$\sim$ symbol, which is more commonly used in the physics-oriented
literature, although less consistent with our source \cite{Dieu}.

\end{document}